\documentclass[11pt]{article}
\usepackage{cite}
\usepackage{graphicx}
\usepackage{rotating}
\usepackage{a4p}
\begin{document}
\newcommand{\etal}{{\it et~al.}}

\begin{center}
{\large EUROPEAN ORGANIZATION FOR NUCLEAR RESEARCH}
\end{center}

\vspace{1cm}
\begin{flushright}
CERN-OPEN-2006-065\\
November 20, 2006
\end{flushright}

\vspace{1.5cm}
\begin{center}
{\large\bf  Quantitative Analysis of the\\ Publishing Landscape in High-Energy Physics}

\vspace{2cm}

Salvatore Mele\footnote{Corresponding author: Salvatore.Mele@cern.ch}$^{,}$\footnote{On leave of absence from INFN,
    Napoli, Italy}, David Dallman,  Jens Vigen, Joanne Yeomans\\ 
CERN, CH-1211, Gen\`eve 23, Switzerland\\

\vspace{2cm}

{\bf  Abstract}
\end{center}
World-wide collaboration in high-energy physics (HEP) is a tradition
which dates back several decades, with scientific publications mostly coauthored
by scientists from different countries. This coauthorship
phenomenon makes it difficult to identify  precisely the ``share'' of
each country in HEP scientific production. One year's worth of HEP
scientific articles published in peer-reviewed journals is analysed
and their authors are uniquely assigned to countries. This
method allows the first correct estimation on a {\it pro rata} basis
of the share of HEP scientific publishing among several countries and
institutions. The results provide an interesting insight into the
geographical collaborative patterns of the HEP community.
The HEP publishing landscape is further analysed to provide
information on the journals favoured by the HEP community and on the
geographical variation of their author bases. These results provide
quantitative input to the ongoing debate on the possible
transition of HEP publishing to an Open Access model.

\newpage

\section{Introduction}

High-energy physics (HEP) is commonly regarded as one of the most
international and collaborative scientific disciplines. Over the last
six decades, large experiments at accelerators of ever-increasing
energy brought together first dozens, then hundreds and now thousands
of scientists from an increasingly wider spectrum of
countries. Furthermore, theoretical HEP predates by a long time
present-day cross-border communication as a truly global
enterprise. This endeavour was fostered by a long-standing tradition
of scientific exchange, regular gatherings and long-term visits to
several major centres of attraction by scientists.

As a consequence of this well-established and thriving cross-border
tradition, coauthorship of HEP articles by scientists affiliated to
institutes in different countries is the norm rather than the
exception. At the same time this coauthorship phenomenon complicates
bibliometric studies aimed at evaluating the relative contributions of
different countries to the production of HEP articles.

This article presents an analysis of the distribution of HEP
authorship over several countries and institutes, taking into account
the coauthorship phenomenon on a {\it pro rata} basis. This analysis
is based on one year's worth of HEP articles, selected as presented in
Section~\ref{sec:2}. Section~\ref{sec:2bis} explains the data-analysis
procedure and discusses some bibliometric results.  Results on the
geographical distribution of HEP authorship are presented in
Section~\ref{sec:3} and then interpreted in Section~\ref{sec:4} in
terms of global collaborative patterns. The publishing landscape is
investigated in Section~\ref{sec:5}, which identifies the journals
most used by HEP authors. Section~\ref{sec:6} presents additional
results on the breakdown of the author base of the leading HEP
journals among different countries; the distribution over different
journals of the HEP scientific production of several countries and
institutes is also discussed.

These results are particularly relevant as they constitute a
quantitative basis for the ongoing debate on the possible transition
of HEP publishing to an Open Access model~\cite{taskForce}. No
assessment of the economical implications of such a transition is
possible without clear and uncontroversial data on the contributions of
different countries to HEP scientific publishing, which is presented
here for the first time.

\section{Data Sample}
\label{sec:2}

The {\it preprint culture} in HEP pioneered the free distribution of scientific results. For
decades, theoretical physicists and scientific collaborations, eager to
disseminate their findings in a way faster than the distribution of
scholarly publications, printed and mailed hundreds, even
thousands, of copies of their manuscripts before submitting them
to peer-reviewed journals. This preprint culture tended, however,
to favour the large laboratories and universities that could afford
mailing large numbers of preprints while receiving comprehensive
regular mailings~\cite{luisella}. The spread of the Internet and the inception of the
{\tt arXiv} repository~\cite{arXiv} ushered a new era for the preprint
culture, offering all scientists a level playing field. In its
current implementation, {\tt arXiv} allows researchers to
submit their preprints and browse or receive regular feeds on recent
submissions in their area of interest~\cite{arxivurl}. The {\tt arXiv} repository
and its mirrors collect the {\it corpus} of HEP articles, classified
into four categories:
\begin{itemize}
\item {\tt hep-ex}, for high energy experimental physics; 
\item {\tt hep-lat} for studies of lattice field theory;
\item {\tt hep-ph} for particle phenomenology;
\item {\tt hep-th} for string, conformal and field theory. 
\end{itemize}
The attribution of articles to a particular category is performed by the
authors themselves at submission time. The system supports cross referencing
while multiple submission is frowned upon so that no double counting of
the same article from two categories is expected in the following analysis.

This analysis is based on all preprints submitted to {\tt arXiv}
in the year 2005 and classified in one of the four HEP
categories. Owing to its widespread preprint culture, this sample
represents a faithful snapshot of HEP peer-reviewed scientific literature.

As in many other disciplines, HEP results are often presented in
preliminary form at international conferences or workshops before
being officially released in the form of a publication in a peer-reviewed
journal. Results are then often summarised at other conferences in
the following years. Preprints usually appear describing these
conference contributions and therefore {\tt arXiv} stores multiple, 
albeit different, entries corresponding to different phases of the
life-cycle of a scientific result.  To avoid this
form of multiple counting of the same piece of work, the following analysis is
restricted to preprints subsequently published in peer-reviewed
journals. This requirement also removes lecture notes, theses and
other unpublished material submitted to {\tt arXiv} but not relevant for this analysis.

The data on which this analysis is based are extracted from the {\tt
SPIRES} database~\cite{spires} hosted at SLAC, the Stanford Linear
Accelerator Center in California, and jointly compiled together with DESY, the
Deutsches Elektronen-Synchrotron in Hamburg, and FNAL, the Fermi
National Accelerator Laboratory in Illinois. This database is chosen
as it has a complete coverage of the HEP articles in {\tt arXiv} and
in addition includes publication information. As an example, the sample
of preprints submitted to the {\tt hep-ex} category in {\tt arXiv}
during 2005, and subsequently published, is obtained with the following
query:
\begin{center} 
{\tt FIND EPRINT HEP-EX/05\# AND PS P AND NOT TYPE C\\
AND NOT TYPE L AND NOT TYPE B AND NOT TYPE T}
\end{center} 
Conference articles, lecture notes, theses and books are
explicitly removed from the search. The samples for the other three
{\tt arXiv} categories are obtained {\it mutatis mutandis}.

\begin{sidewaystable}[p]
  \begin{center}
    \begin{tabular}{|c|r|r|c|r|r|c|r|r|c|r|r|c||r|r|c|}
      \cline{2-16}
      \multicolumn{1}{c|}{}&\multicolumn{3}{c|}{hep-ex}&\multicolumn{3}{c|}{hep-lat}&\multicolumn{3}{c|}{hep-ph}&\multicolumn{3}{c||}{hep-th}&\multicolumn{3}{c|}{Total}\\
      \hline
      Year& $N_S$ & $N_P$ & $\varepsilon$ & $N_S$ & $N_P$ & $\varepsilon$ & $N_S$ & $N_P$ & $\varepsilon$ & $N_S$ & $N_P$ & $\varepsilon$ & $N_S$ & $N_P$ & $\varepsilon$ \\
      \hline
      2005 & 854 & 338 & 40\% & 663 & 246 & 37\% & 3918 & 2207 & 56\% & 3238 & 2225 & 69\% & 8673 & 5016 & 58\%\\
      \hline
      2004 & 885 & 349 & 39\% & 586 & 261 & 45\% & 4138 & 2534 & 61\% & 3357 & 2361 & 70\% & 8966 & 5505 & 61\%\\
      2003 & 771 & 287 & 37\% & 575 & 227 & 39\% & 3964 & 2381 & 60\% & 3275 & 2428 & 74\% & 8585 & 5323 & 62\%\\
      2002 & 885 & 293 & 33\% & 583 & 218 & 37\% & 4245 & 2383 & 56\% & 3333 & 2482 & 74\% & 9046 & 5376 & 59\%\\
      2001 & 819 & 328 & 40\% & 574 & 218 & 38\% & 4228 & 2499 & 59\% & 3181 & 2305 & 72\% & 8802 & 5350 & 61\%\\
      2000 & 735 & 324 & 44\% & 508 & 235 & 46\% & 4124 & 2390 & 58\% & 3144 & 2259 & 72\% & 8511 & 5208 & 61\%\\
      1999 & 666 & 317 & 48\% & 588 & 244 & 41\% & 4076 & 2602 & 64\% & 2825 & 2180 & 77\% & 8155 & 5343 & 66\%\\
      1998 & 406 & 231 & 57\% & 623 & 282 & 45\% & 3807 & 2442 & 64\% & 2774 & 2061 & 74\% & 7610 & 5016 & 66\%\\
      1997 & 325 & 192 & 59\% & 548 & 227 & 41\% & 3615 & 2305 & 64\% & 2865 & 1990 & 69\% & 7353 & 4714 & 64\%\\
      1996 & 166 & 82 & 49\% & 558 & 248 & 44\% & 3327 & 2149 & 65\% & 2626 & 1924 & 73\% & 6677 & 4403 & 66\%\\
      1995 & 158 & 99 & 63\% & 437 & 228 & 52\% & 2990 & 2008 & 67\% & 2347 & 1704 & 73\% & 5932 & 4039 & 68\%\\
      1994 & 67 & 35 & 52\% & 447 & 202 & 45\% & 2500 & 1714 & 69\% & 2349 & 1639 & 70\% & 5363 & 3590 & 67\%\\
      1993 & $-$ & $-$ &$-$ & 374 & 209 & 56\% & 1762 & 1275 & 72\% & 2084 & 1460 & 70\% & 4220 & 2944 & 70\%\\
      1992 & $-$ & $-$ &$-$ & 321 & 180 & 56\% & 755 & 559 & 74\% & 1378 & 1080 & 78\% & 2454 & 1819 & 74\%\\
      1991 & $-$ & $-$ &$-$ & 4 & 3 & 75\% & $-$ & $-$ & $-$ & 302 & 228 & 75\% & 306 & 231 & 75\%\\
      \hline
    \end{tabular}
    \caption{Numbers of preprints submitted to the different {\tt
arXiv} HEP categories ($N_S$) and subsequently published in
peer-reviewed journals ($N_P$) together with their total. The ratio
$\varepsilon=N_S/N_P$ is also listed. Figures are given for the entire
{\tt arXiv} historical sample. Data corresponding to the year 2005 is
used in this analysis.}
  \end{center}
  \label{tab:0}
\end{sidewaystable}

\section{Data Analysis}
\label{sec:2bis}

Table~1 presents the numbers of hits obtained by the {\tt SPIRES} query in
the four categories and their sum for the year 2005 as well as the
entire historical record. A total of 5016 articles are selected for
the year 2005. The total numbers of submissions for each {\tt arXiv} category
obtained with queries such as: 
\begin{center} 
{\tt FIND EPRINT HEP-EX/05\#}
\end{center}
are also presented in Table~1 together with their
sum. The difference with the sample considered in this article is
composed of conference articles and unpublished material. The ratios
of the numbers of published articles to the numbers of {\tt arXiv}
submissions is also presented in Table~1. 

The historical evolution of the numbers in Table~1 is
interesting: early years show a gradual increase in the number of
submissions, consistent with the gradual adoption of the system, while
numbers for later years are consistent with a plateau structure with
year-to-year variations of a few percentage points.

The queries on which this article is based were performed in the
second half of October 2006 and one could argue that some preprints
submitted in late 2005 could have still been in the editorial process and
would not therefore have yet appeared in peer-reviewed journals. If the
five-year period $2000-2004$ is used to predict the number of articles
extracted by the query for the year 2005, this is just 6\% above the
number actually observed, leading to the conclusion that no large
systematic bias affects the size of the sample under
consideration. There are no reasons to believe that any sizable
systematic effect from a small fraction of ``undiscovered'' articles
would affect the relative contributions of different countries 
presented in the following.

Figure~1 presents the distribution among the four different
{\tt arXiv} categories of the 5016 articles on which this analysis is
based. Experimental results account for just 6.7\% of the total.

\begin{figure}[h]
  \begin{center}
    \includegraphics[width=\textwidth]{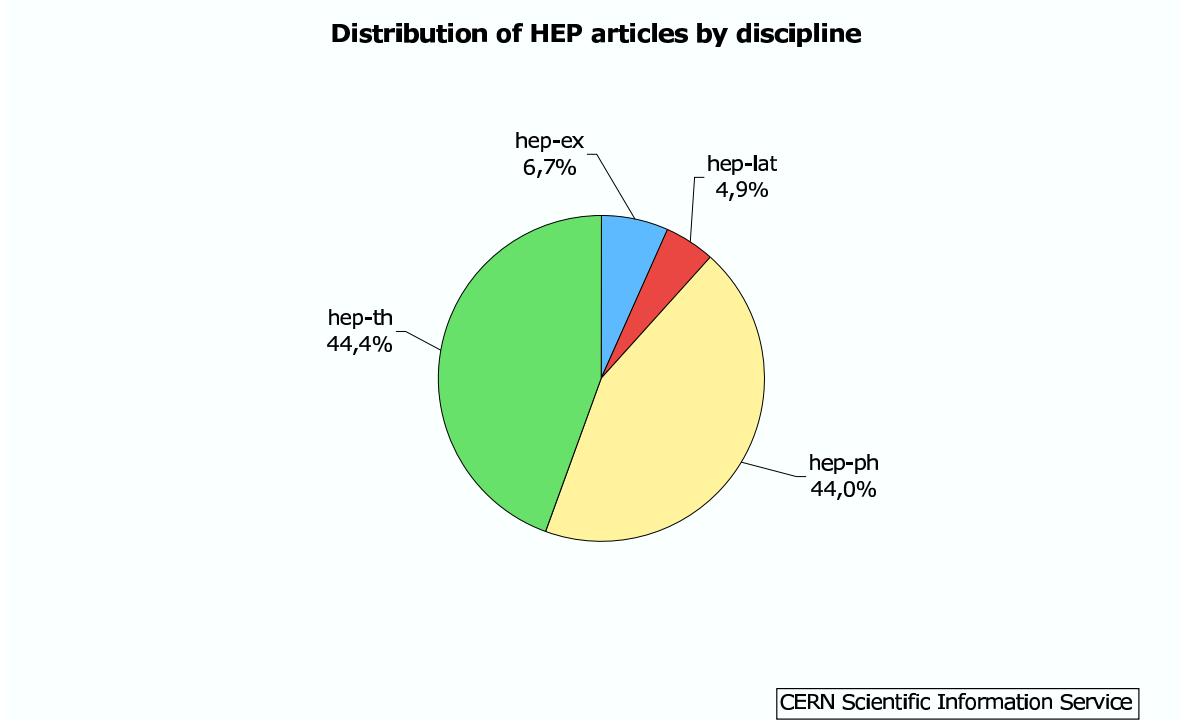}
    \caption{Distribution by {\tt arXiv} category of the sample used
      in this analysis, corresponding to 5016
      preprints submitted in the year 2005 and subsequently published
      in peer-reviewed journals.}
  \end{center}
  \label{fig:0}
\end{figure}\clearpage

\begin{figure}[h]
  \begin{center}
    \includegraphics[width=0.75\textwidth]{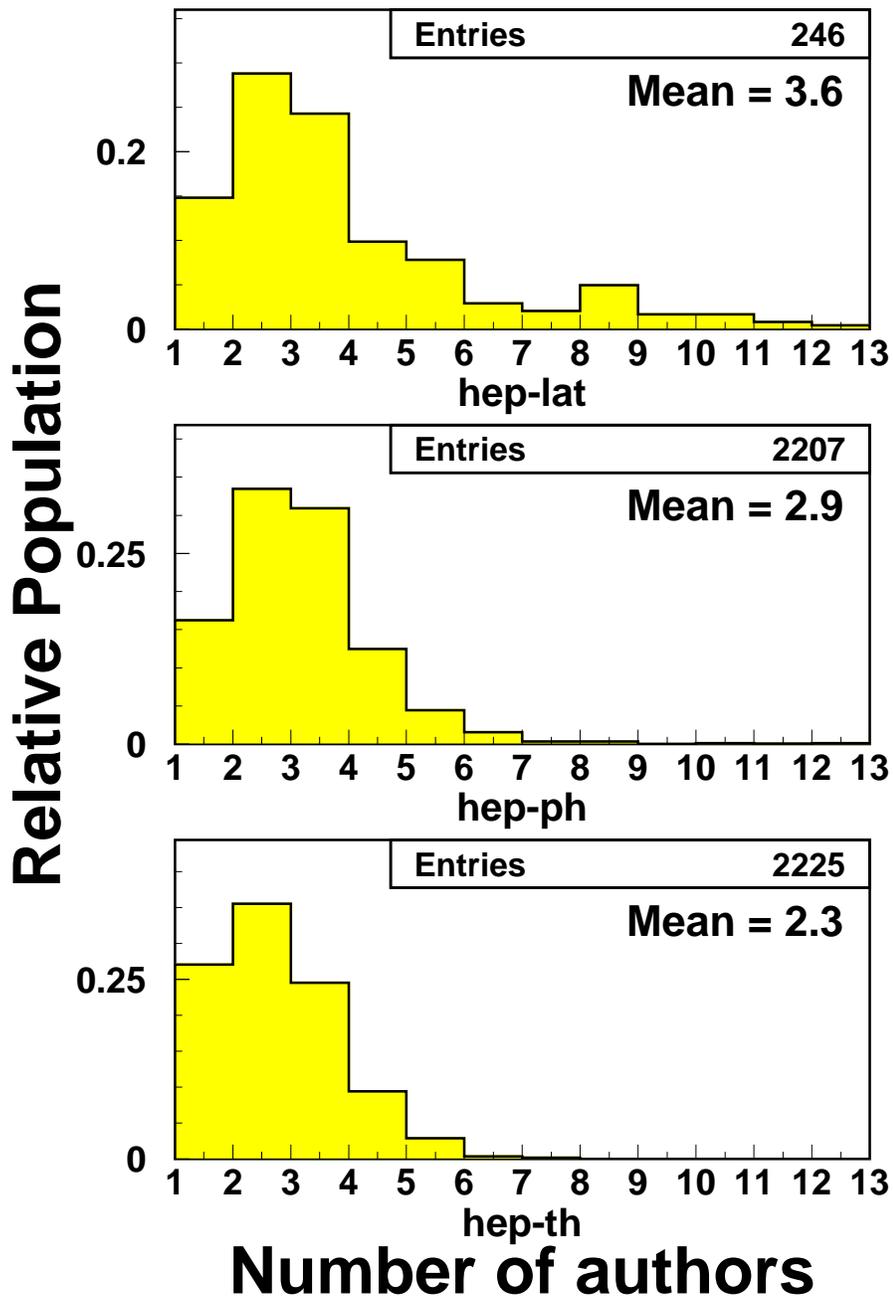}
    \caption{Distributions of the number of authors of {\tt hep-lat},
{\tt hep-ph} and {\tt hep-th} articles used in this analysis. 
The distributions are normalised to unit area and their
mean is indicated.}
  \end{center}
  \label{fig:1}
\end{figure}\clearpage

A first bibliometric result extracted from this study is the
distribution of the number of authors per article. Figure~2
presents the distribution of the number of authors of each article in
the three non-experimental classes {\tt hep-lat}, {\tt hep-ph} and
{\tt hep-th}. The average number of authors for the three classes are
3.6, 2.9 and 2.3, respectively. The average number of authors for the
sum of the three classes is 2.6. The average number of authors for the
{\tt hep-ex} class is about 290. The distribution of the number of
authors is biased by the fact that a dozen large experimental
collaborations appear several times in the data sample. The breakdown
of the considered {\tt arXiv:hep-ex} sample into different experiments
is  shown in Figure~3. Implications of the large number of authors in
experimental collaborations are discussed in Reference~\cite{iupap}.

\begin{figure}[h]
  \begin{center}
    \includegraphics[width=\textwidth]{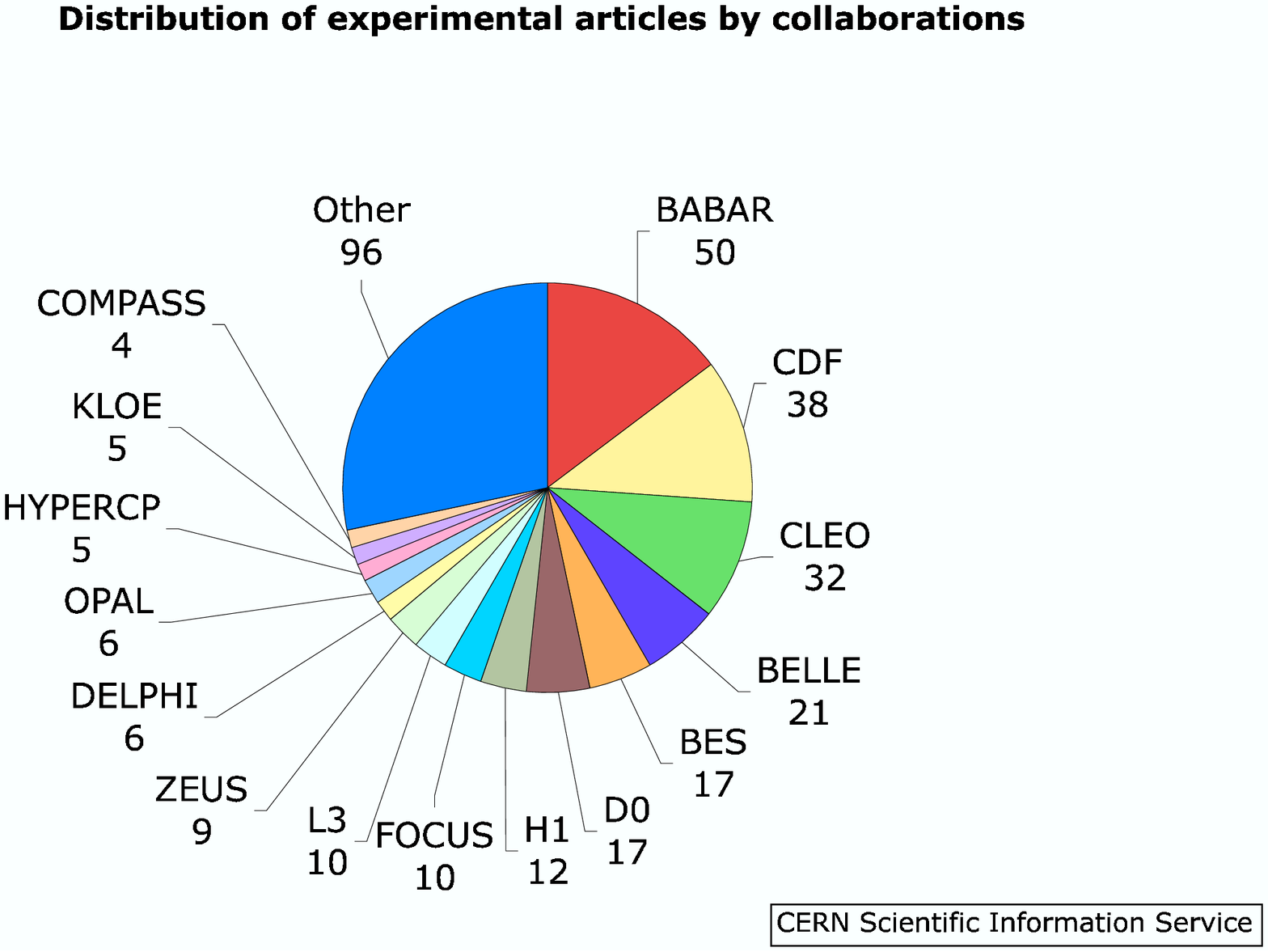}
    \caption{Number of articles from the large experimental collaborations
      submitted to {\tt
        arXiv:hep-ex} in 2005 and subsequently published in peer-reviewed
      journals. The ``Other'' category comprises collaborations which
      published
      less than 4 articles as well as articles with less than 40
      authors. The total number of articles is 338.}
  \end{center}
  \label{fig:coll}
\end{figure}\clearpage

Unfortunately, as of today, no database allows an automatic extraction
of bibliographic information concerning author affiliations for HEP
articles at the level needed for this analysis. Therefore each article
satisfying the query had to be inspected to perform a manual classification of the authors according to their
affiliation. The output format of {\tt SPIRES} partly alleviates this
problem as author affiliations are often readable off the standard
web-based output of the queries without having to access the article
metadata on a publisher's web site or the full-text version in {\tt
  arXiv}. Author affiliations were classified into 22 classes, listed
in the first column of Table~2.  European, American
and Asian countries are singled out according to their contribution to
the global HEP scientific production, down to a lower limit of about
1\%. The contribution from CERN, the world's largest HEP laboratory, is
shown separately. The remaining countries are divided into two classes:
CERN Member 
States\footnote{CERN Member States not already listed in the first
  column of Table~2 are: Austria, Belgium, Bulgaria, the
  Czech Republic, Denmark, Finland, Greece, Hungary,  Norway, Poland
  and the Slovak Republic.} 
and the remaining countries. As the vast majority of HEP in Italy is
funded by INFN, the Istituto Nazionale di Fisica Nucleare, its
contribution has been considered {\it in lieu} of the Italian
one. Italian authors without an INFN affiliation are counted in the
``Other Member States'' category.

As mentioned above, medium- and long-term visits of authors to
different institutes and major laboratories is the staple diet of the
HEP collaborative soul. As a consequence, authors of HEP articles
often have multiple affiliations. Three principles to assign authors
with multiple affiliations to a single class are followed in the order
they are presented below.
\begin{enumerate}
\item If one of the multiple affiliations of an author is a HEP
laboratory, the author is assigned to that laboratory in the case
of CERN, or to the host nation of the laboratory in the other cases.
\item If only one of the multiple affiliations of an author
corresponds to one of the countries explicitly singled out for the
analysis, the author is assigned to that country.
\item If more than one of the multiple affiliations of an author
corresponds to one of the countries explicitly singled out for the
analysis, the author is assigned to a country or institution, 
according to an indicator which takes into account their {\it
pro-capita} Gross Domestic Product and their expected share of the HEP
scientific production.
\end{enumerate}

\section{Distribution of the HEP Production by Country}
\label{sec:3}

The first result of this analysis is the calculation of the share of
HEP publications authored by each of the 22 countries and institutions
into which the authors are classified. For each article in one of the
four {\tt arXiv} categories, each of the 22 countries and institutions
is attributed a fraction of the article corresponding to the number of
authors associated to that country, divided by the total number of
authors. The sum of these fractions over all the articles of an {\tt
arXiv} category, divided by the total number of articles in that
category, defines the share of a particular country or
institution. The results are listed in Table~2 for the four
{\tt arXiv} categories as well as for their
average. Figure~4 presents the
distribution of the HEP scientific production over different
countries. To our knowledge, this is the first result on the
distribution of the HEP scientific literature by country where the
phenomenon of coauthorship is taken into account.

It is interesting to combine the results presented in
Table~2 into the three largest sections of  HEP authorship:
CERN and its Member States, the United States, and the remaining
countries. These results are presented in Table~3 for the
four {\tt arXiv} classes and their average. Figures~5
and~6 show a summary of the distributions of HEP
authorship for the {\tt arXiv} classes and their average,
respectively.

\begin{table}[hbt]
  \begin{center}
    \begin{tabular}{|l|r|r|r|r|r|}
      \hline
      &                    hep-ex& hep-lat& hep-ph& hep-th&  Average\\  
      \hline  
      CERN&                    0.9\%&  1.1\%&   1.7\%&  1.1\%&    1.3\%\\             
      Germany&                 6.3\%&  19.5\%&  10.3\%& 6.5\%&    8.8\%\\             
      UK&                      6.4\%&  6.3\%&   6.6\%&  8.5\%&    7.4\%\\             
      INFN&                    11.0\%& 5.8\%&   5.6\%&  5.3\%&    5.8\%\\             
      France&                  4.1\%&  2.0\%&   3.3\%&  3.2\%&    3.2\%\\             
      Spain&                   0.8\%&  1.2\%&   3.5\%&  2.6\%&    2.8\%\\             
      Switzerland&             1.2\%&  1.1\%&   1.2\%&  0.9\%&    1.1\%\\             
      Sweden&                  0.2\%&  1.2\%&   0.8\%&  1.0\%&    0.9\%\\             
      Portugal&                0.3\%&  0.5\%&   1.4\%&  0.5\%&    0.9\%\\             
      Netherlands&             0.6\%&  0.5\%&   0.5\%&  1.4\%&    0.9\%\\             
      Other Member States&     3.5\%&  3.3\%&   6.7\%&  7.9\%&    6.8\%\\             
      Russia&                  5.1\%&  3.5\%&   5.6\%&  4.0\%&    4.8\%\\             
      Israel&                  0.3\%&  0.8\%&   0.9\%&  1.3\%&    1.0\%\\             
      United States&          40.2\%& 30.0\%&  22.8\%& 22.3\%&   24.1\%\\            
      Canada&                  1.8\%&  1.7\%&   2.0\%&  3.6\%&    2.7\%\\             
      Brazil&                  0.7\%&  0.8\%&   1.9\%&  3.8\%&    2.6\%\\             
      India&                   0.4\%&  2.0\%&   2.7\%&  3.0\%&    2.6\%\\             
      Japan&                   6.3\%&  9.2\%&   6.4\%&  8.4\%&    7.4\%\\             
      China&                   6.4\%&  2.3\%&   6.6\%&  2.6\%&    4.6\%\\             
      Korea&                   1.1\%&  0.2\%&   1.8\%&  2.0\%&    1.8\%\\             
      Taiwan&                  1.1\%&  0.5\%&   1.6\%&  0.8\%&    1.2\%\\             
      Other Countries&         1.1\%&  6.5\%&   6.0\%&  9.3\%&    7.2\%\\
      \hline
    \end{tabular}
    \caption{Distribution of HEP scientific
literature over different countries and institutions for the four HEP {\tt
arXiv} classes and their average.}
  \end{center}
  \label{tab:1}
\end{table}

\begin{table}[h!]
  \begin{center}
    \begin{tabular}{|l|r|r|r|r|r|}
      \hline
      &hep-ex&hep-lat&hep-ph&hep-th&Average\\
      \hline
      CERN \& Member States&35.5\%&42.3\%&41.6\%&38.8\%&40.0\%\\
      United States&40.2\%&30.0\%&22.8\%&22.3\%&24.1\%\\
      Other Countries&24.3\%&27.7\%&35.6\%&38.9\%&35.9\%\\
      \hline
    \end{tabular}
    \caption{Distribution of HEP scientific production over three
      geographical groups for the four HEP {\tt
arXiv} classes and their average.}
  \end{center}
  \label{tab:2}
\end{table}

\begin{sidewaysfigure}[p]
  \begin{center}
    \includegraphics[width=\textheight]{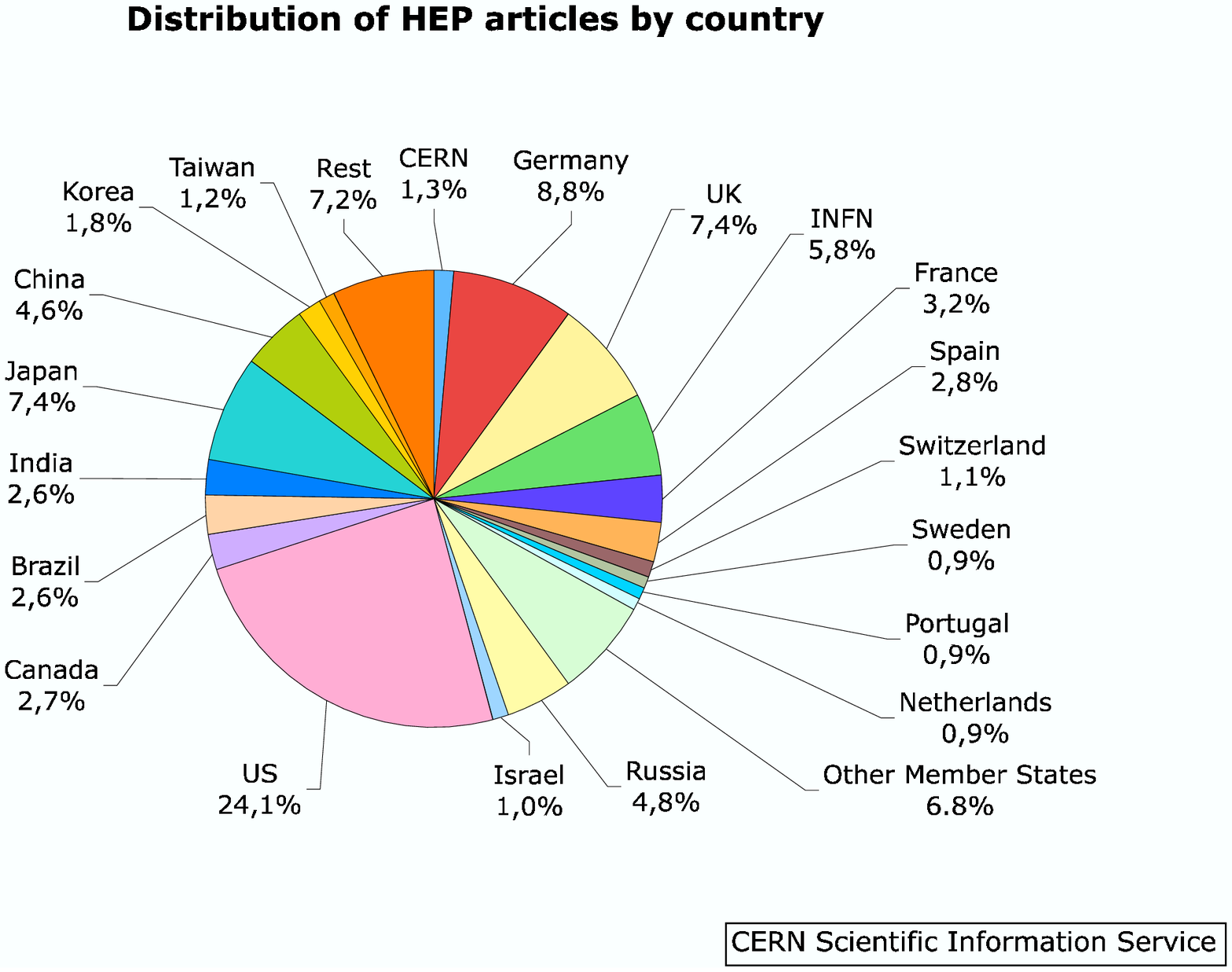}
    \caption{Distribution of the HEP scientific
literature over different countries and institutions. A sample of 5016
articles submitted to {\tt arXiv} in 2005 and subsequently published
in peer-reviewed journals is considered. Coauthorship is taken into
account by assigning fractions of articles to different countries on a
{\it pro-rata} basis.}
  \end{center}
  \label{fig:2c}
\end{sidewaysfigure}

\begin{figure}[t]
  \begin{center}
    \begin{tabular}{cc}
      \mbox{\includegraphics[width=0.5\textwidth]{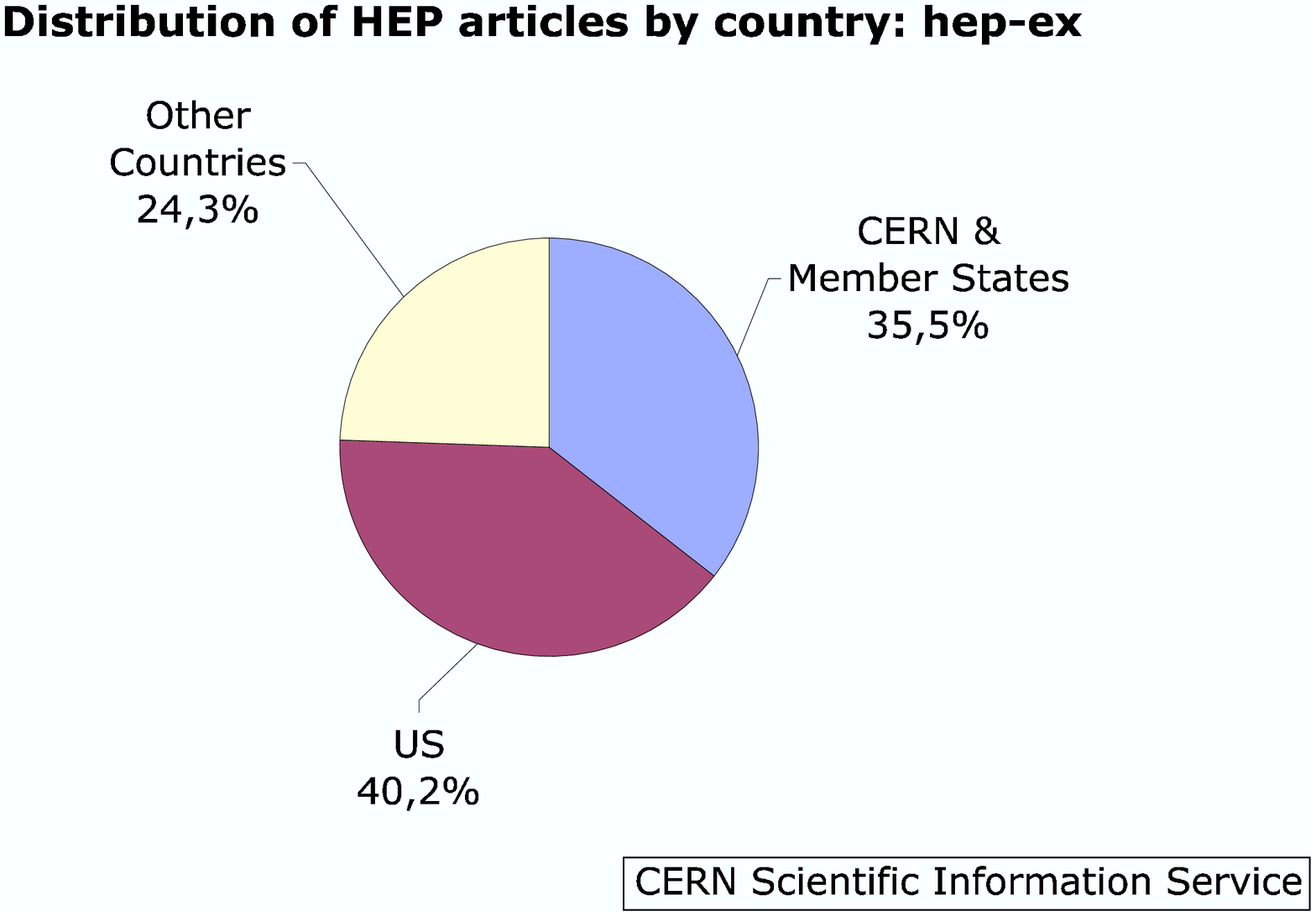}}&
      \mbox{\includegraphics[width=0.5\textwidth]{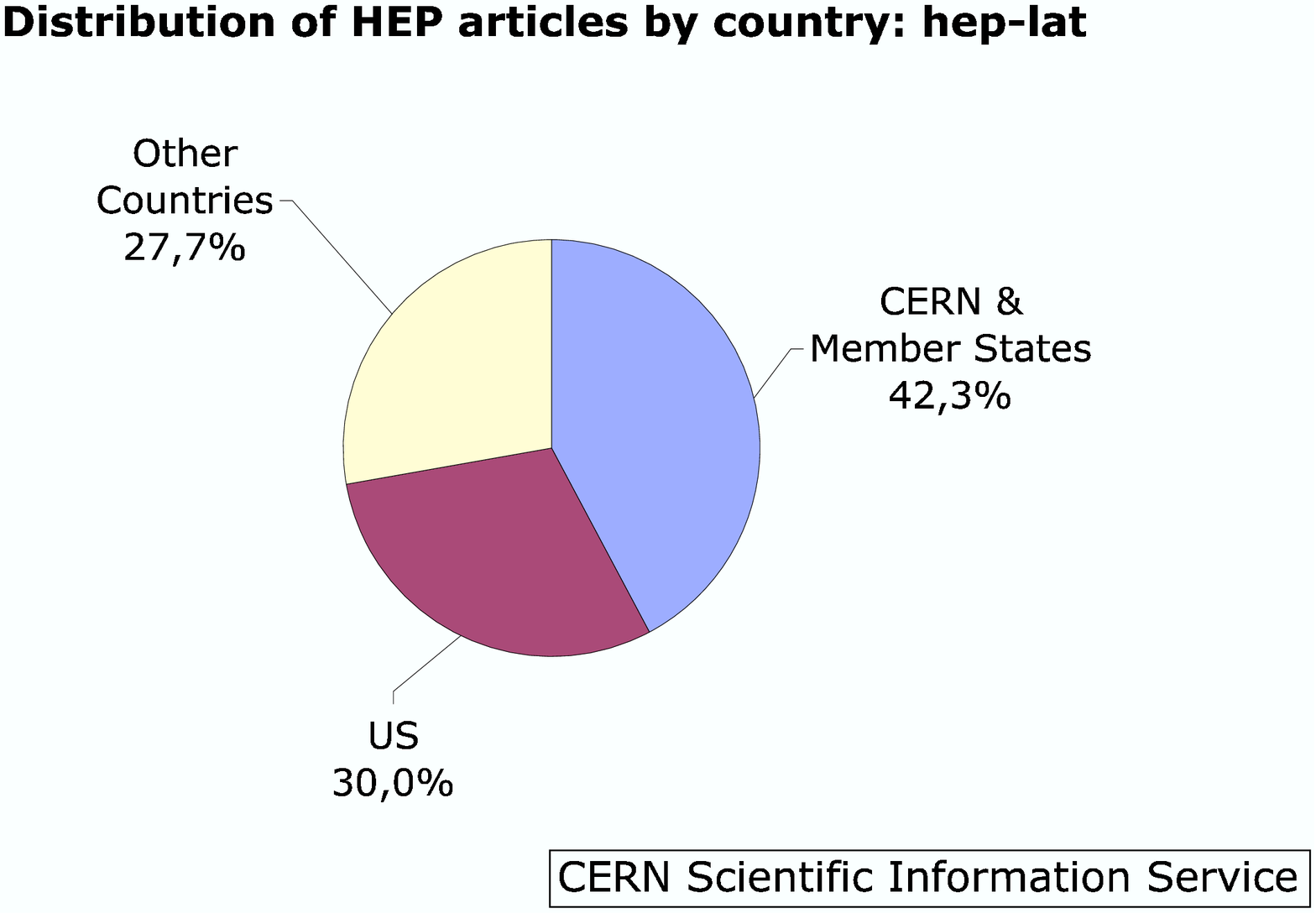}}\\
      \mbox{\includegraphics[width=0.5\textwidth]{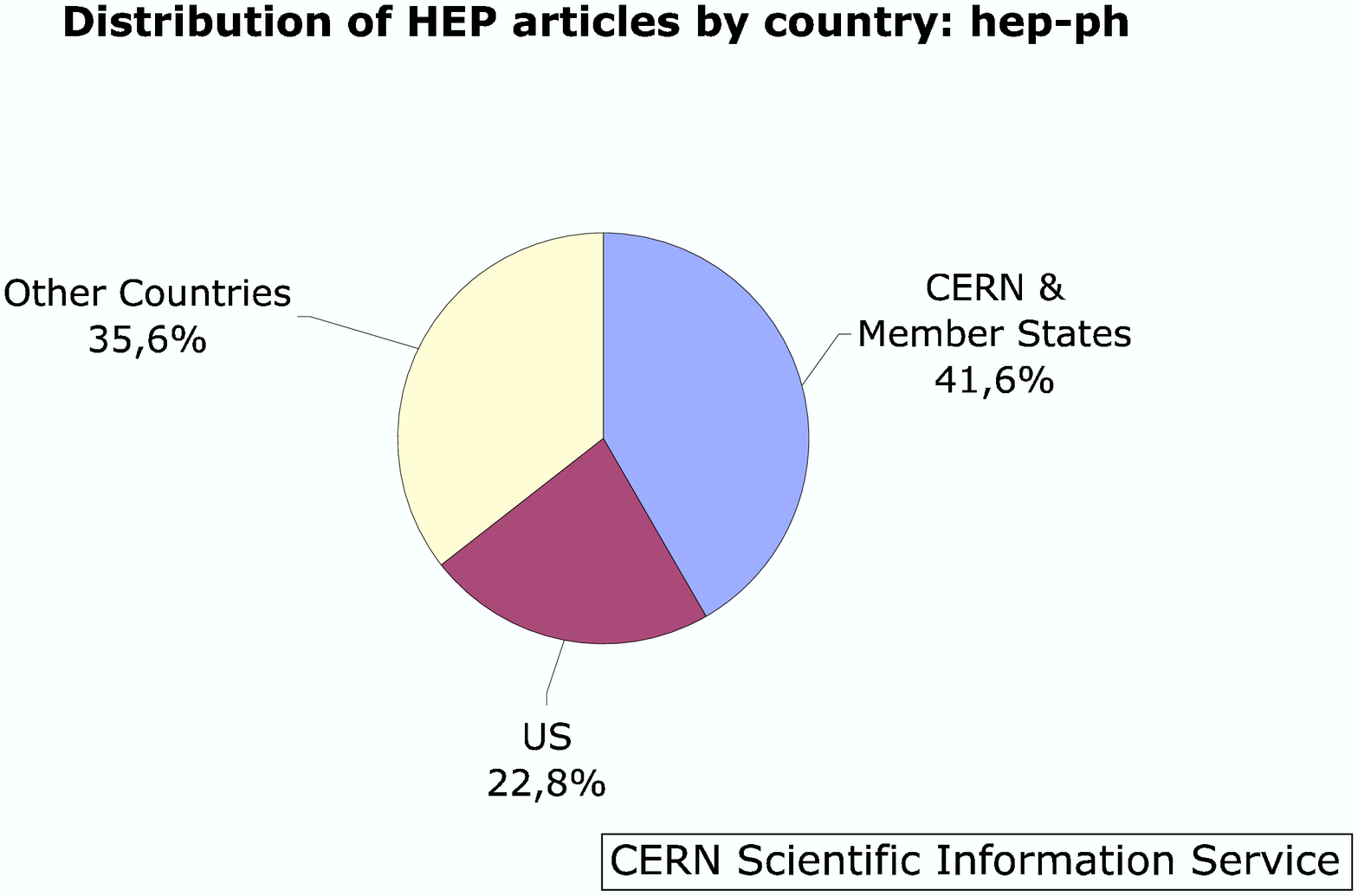}}&
      \mbox{\includegraphics[width=0.5\textwidth]{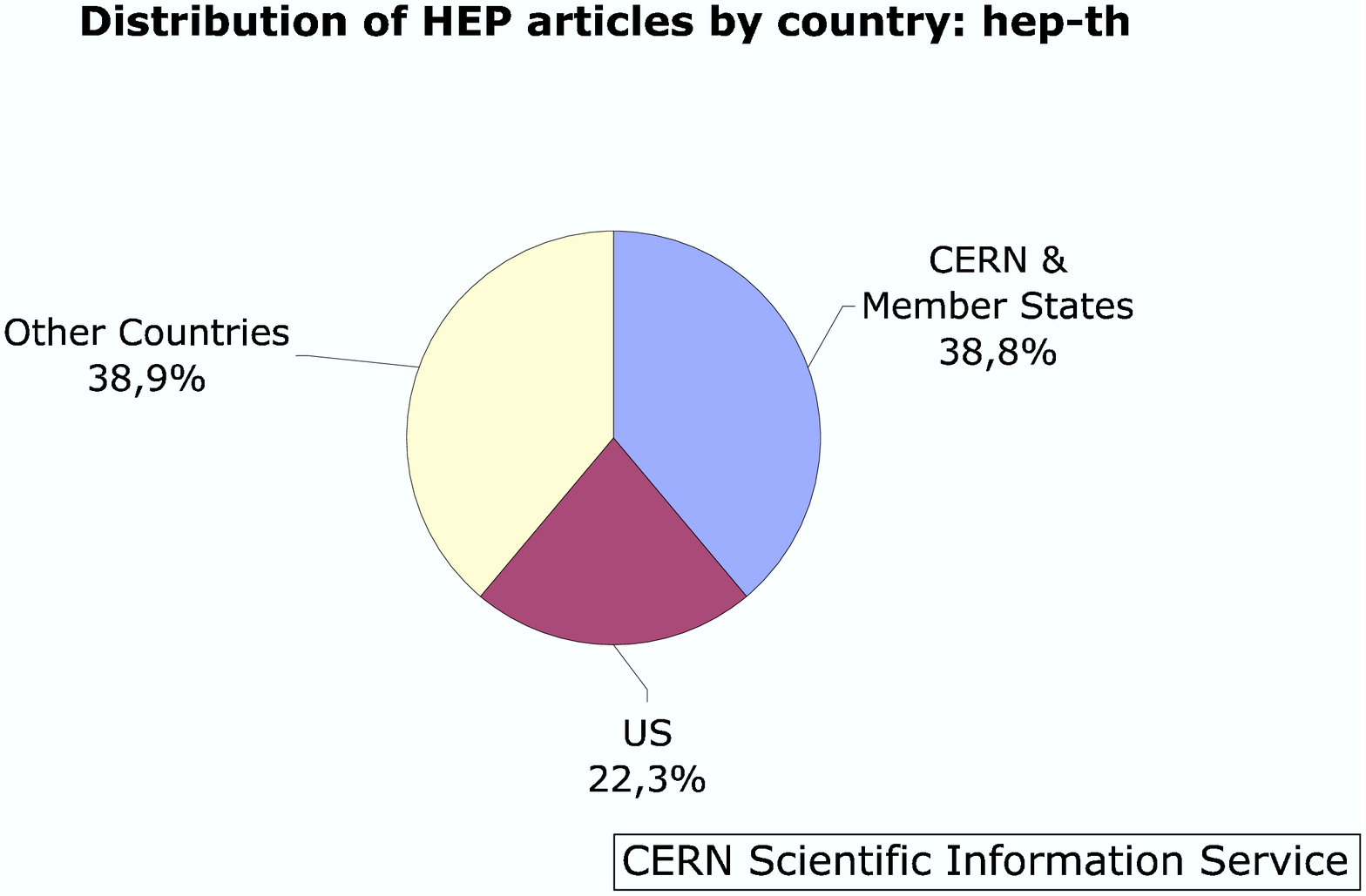}}\\
      \end{tabular}
    \caption{Distribution of HEP scientific production over three
      geographical groups for the four {\tt
arXiv} classes.}
  \end{center}
  \label{fig:2b}
\end{figure}

\begin{figure}[h!]
  \begin{center}
    \includegraphics[width=\textwidth]{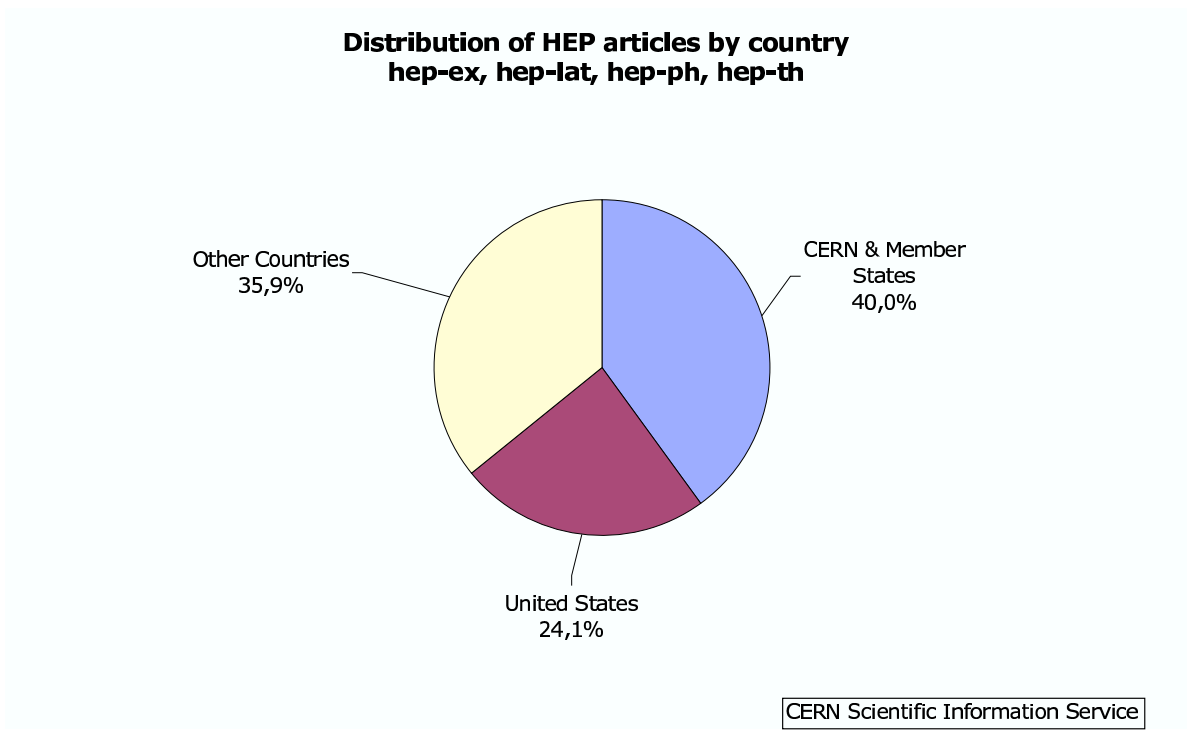}
    \caption{Distribution of HEP scientific production over three
      geographical groups.}
  \end{center}
  \label{fig:2a}
\end{figure}\clearpage

\section{Collaborative Patterns in HEP}
\label{sec:4}

The data sample under investigation allows a study of the
collaborative patterns in HEP in order to answer a natural question:
which groups of countries and institutions collaborate? A
simplified approach to address this question is chosen, in which only
three large groups of authors are considered, according to their
affiliation to one of three sections of HEP authorship: CERN and its
Member States, the United States, and the remaining countries. Results
from more complex analyses of other data samples focusing on author-to-author collaborative
networks are presented in Reference~\cite{colla}. 
Each article is assigned to one of seven mutually-exclusive classes:
\begin{enumerate}
\item all the authors are associated to CERN or any of its Member States;
\item all the authors are associated to the United States;
\item no authors are associated to CERN, its Member States or the
  United States;
\item some authors are associated to CERN or one of its Member States and
  some to the United States, but none to any other country;
\item some authors are associated to CERN or one of its Member States and
  some to other countries, but none to the United States;
\item some authors are associated to the United States  and
  some to other countries but none to CERN or any of its Member States;
\item at least one author is associated to CERN or one of its Member States,
  one to the United States and one to some other country.
\end{enumerate}
Figure~7 presents the fraction of HEP articles in each of
these seven classes while
Figure~8 shows the results for the four
separate {\tt arXiv} disciplines.

\section{Distribution of HEP Publications among Journals}
\label{sec:5}

The 5016 articles considered in this study appeared in 89 different
peer-reviewed journals. The distribution of articles over the
different journals is presented in Table~4 for the four different HEP
disciplines and their global average, which is also shown in
Figure~9. Only the 11 journals with a share above 1\% are considered
in Table~4 and Figure~9. However, the share of Nuclear Instruments and
Methods in Physics Research (NIM) is also singled out. The
contribution to this journal is interesting as this title is the
reference journal for instrumentation in HEP. The low share of this
journal in the total is due to the reduced contribution of
experimental HEP to the total production compared to the theoretical
and phenomenological studies, as presented in Figure~1. However, the
low percentage of instrumentation articles among the total amount of
experimental articles, 2.7\%, is also due to the far less widespread
culture of self-archiving results in {\tt arXiv} in the HEP
instrumentation community. A direct inspection of articles published
in NIM in 2005 revealed about 30\% of articles of potential interest
for HEP instrumentation which had not been submitted to {\tt arXiv},
neither before nor after publication.

\begin{figure}[p]
  \begin{center}
      \mbox{\includegraphics[width=0.7\textwidth]{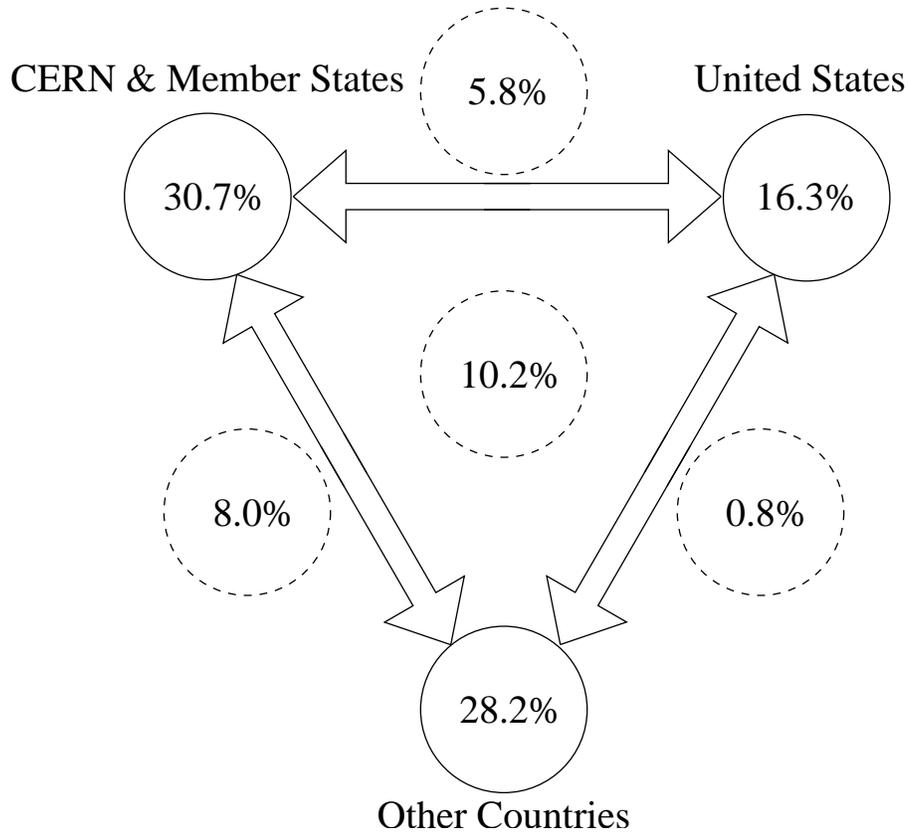}}
    \caption{Collaborative patterns in HEP. Numbers in the circles at
      the vertices of the triangle represent the percentages of articles
      produced by individual authors or authors collaborating with others within the same
      group of countries and institutions. Numbers in the dashed circles
      along the sides of the triangle represent the percentages of articles
      produced by collaborations of authors from countries and institutions
      in the two groups indicated by the neighbouring vertices. The number in the
      dashed circle in the centre of the triangle represents the articles
      produced by collaborations spanning the three groups of countries. The
      plot presents results for the entire HEP production submitted
      to {\tt arXiv} in 2005 and subsequently published.}
  \end{center}
  \label{fig:3a}
\end{figure}\clearpage

\begin{figure}[p]
  \begin{center}
      \mbox{\includegraphics[width=0.7\textwidth]{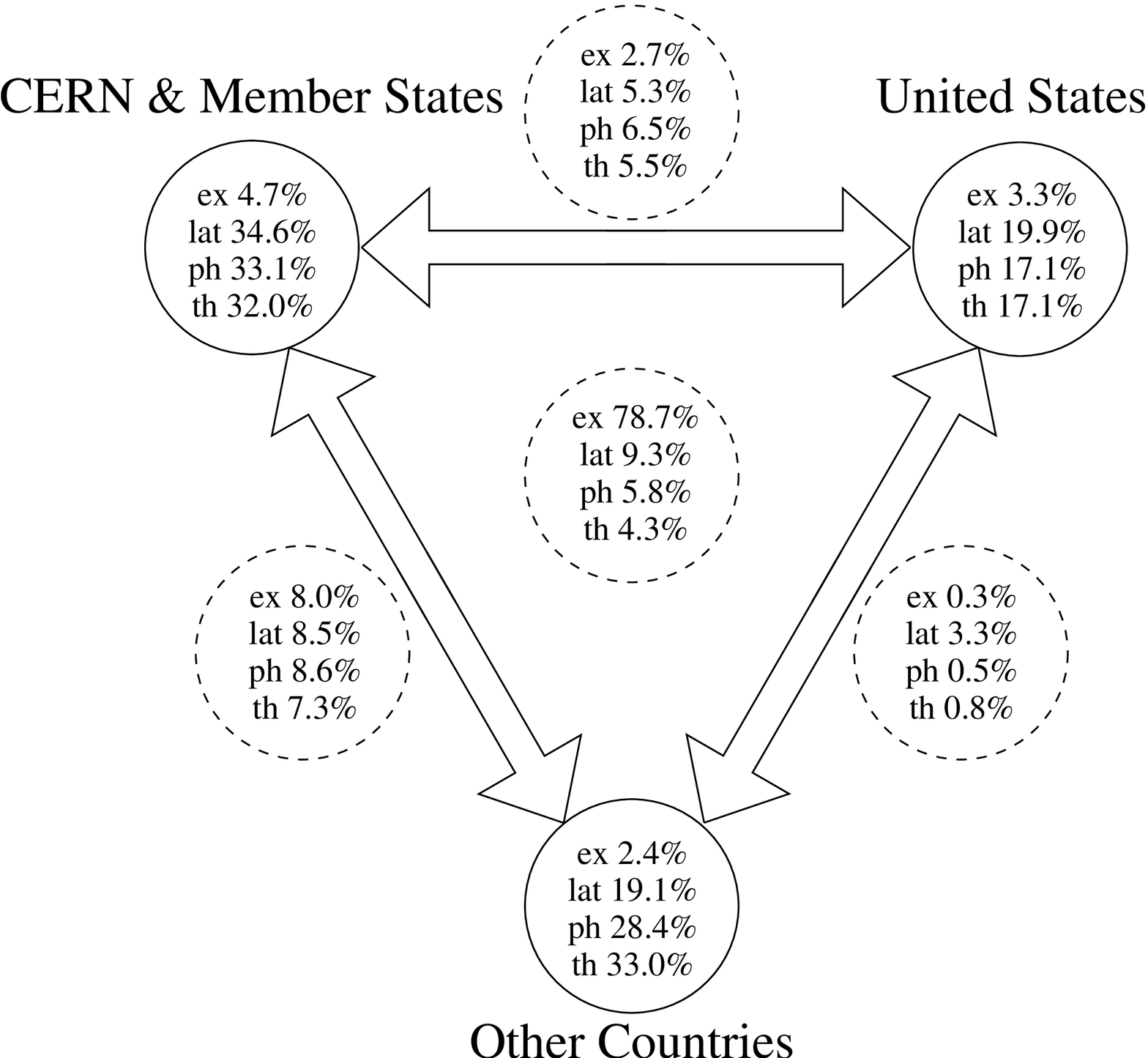}}
    \caption{Collaborative patterns in HEP. Collaborative patterns in HEP. Numbers in the circles at
      the vertices of the triangle represent the percentages of articles
      produced by individual authors or authors collaborating with others within the same
      group of countries and institutions. Numbers in the dashed circles
      along the sides of the triangle represent the percentages of articles
      produced by collaborations of authors from countries and institutions
      in the two groups indicated by the neighbouring vertices. The numbers in the
      dashed circle in the centre of the triangle represents the articles
      produced by collaborations spanning the three groups of countries. The
      plot presents the results for each of the four disciplines in which
      {\tt arXiv} preprints are classified by the authors.}
  \end{center}
  \label{fig:3b}
\end{figure}\clearpage

\begin{sidewaysfigure}[p]
  \begin{center}
    \includegraphics[width=\textheight]{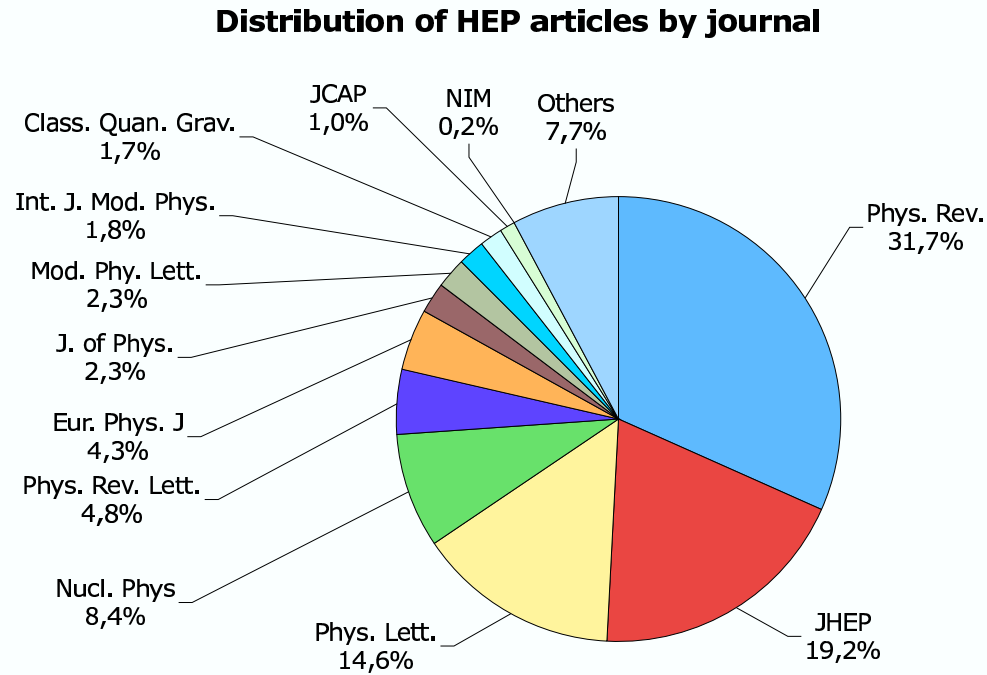}
    \caption{Distribution of the HEP articles in several journals. Only
      journals with a total share above 1\% are considered, with the
      exception of Nuclear Instruments and Methods in Physics Research
      (NIM). The remaining 77 journals are grouped under
      ``Others''. Journals are ordered clockwise according to decreasing
      shares. A total of 83\% of HEP articles is published in
      just six journals.}
  \end{center}
  \label{fig:4}
\end{sidewaysfigure}\clearpage

An analysis of the results in Table~4 shows that 83\% of HEP articles
are published in just six journals: Physical Review (A through
E); Journal of High Energy Physics (JHEP); Physics Letters (A and B);
Nuclear Physics (A and B); Physical Review Letters and the European
Physical Journal (A and C).

\begin{table}[h!]
  \begin{center}
    \begin{tabular}{|l|l|r|r|r|r|r|}
      \hline
      Journal&Publisher&hep-ex&hep-lat&hep-ph&hep-th&Average\\
      \hline
      Phys. Rev.&APS&31.7\%&52.8\%&41.5\%&19.7\%&31.7\%\\
      JHEP&SISSA&$-$&14.2\%&10.0\%&31.8\%&19.2\%\\
      Phys. Lett.&Elsevier&21.3\%&15.9\%&16.4\%&11.6\%&14.6\%\\
      Nucl. Phys.&Elsevier&1.2\%&6.5\%&7.3\%&10.7\%&8.4\%\\
      Phys. Rev. Lett.&APS&29.0\%&2.4\%&4.4\%&1.8\%&4.8\%\\
      Eur. Phys. J.&Springer&10.7\%&2.0\%&7.0\%&1.0\%&4.3\%\\
      J. of Phys.&IOP&$-$&0.8\%&2.1\%&3.1\%&2.3\%\\
      Mod. Phys. Lett.&World Scientific&1.2\%&0.8\%&2.3\%&2.6\%&2.3\%\\
      Int. J. Mod. Phys.&World Scientific&0.3\%&1.6\%&1.4\%&2.3\%&1.8\%\\
      Class. Quan. Grav. &IOP&$-$&$-$&0.1\%&3.8\%&1.7\%\\
      JCAP&SISSA&$-$&$-$&1.0\%&1.3\%&1.0\%\\
      NIM&Elsevier&2.7\%&$-$&0.1\%&$-$&0.2\%\\
      Others&$-$&2.1\%&2.8\%&6.5\%&10.2\%&7.7\%\\
      \hline
    \end{tabular}
    \caption{Distribution of HEP articles over different journals
      for the four HEP {\tt arXiv} classes and their average. Only
      journals with a total share above 1\% are considered, with the
      exception of Nuclear Instruments and Methods in Physics Research
      (NIM). The remaining 77 journals are grouped under
      ``Others''. The publishers of the different journals are also indicated.}
  \end{center}
  \label{tab:3}
\end{table}

These six journals are published by just four publishers: the American Physical Society,
Elsevier, SISSA and Springer, as detailed in  Table~4. It is interesting to split the corpus of HEP
scientific literature discussed in this article according to the
publisher of the journal in which the article appeared. The results
are presented in Figure~10. A total of 87\% of HEP articles are
published by the same four publishers listed above.

\begin{figure}[tb]
  \begin{center}
    \includegraphics[width=0.9\textwidth]{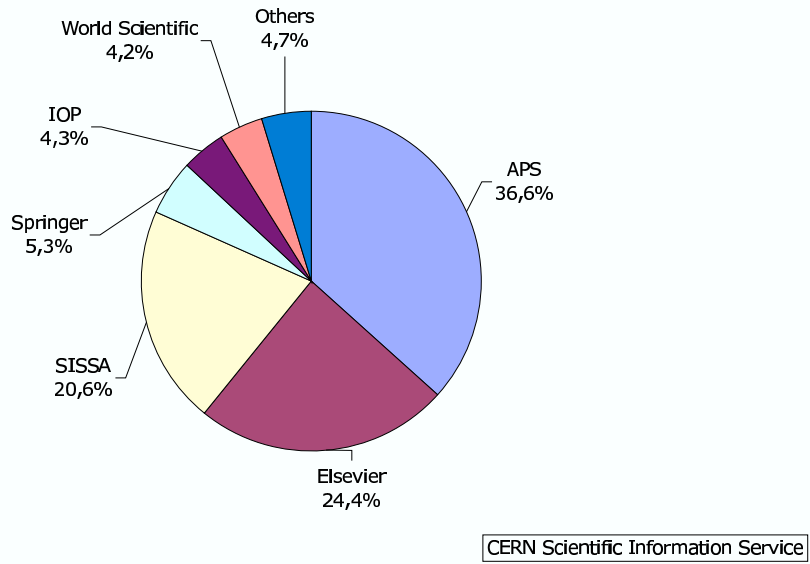}
    \caption{Distribution of HEP articles over different
      publishers. A total of 87\% of  HEP articles are published by
      four publishers: APS, Elsevier, SISSA and Springer.}
  \end{center}
  \label{fig:5}
\end{figure}

\section{Geographical Analysis of HEP Journals}
\label{sec:6}

The quantitative information on the different countries and
institutions contributing to each of the HEP articles considered in
this analysis allows the estimation of the geographical distribution of the
authors for each of the 12 journals listed in
Table~4. The analysis of Section~\ref{sec:3} is repeated for
each journal and the results are presented in Table~5 for
all 22 countries and institutions considered in this article, as well
as their grouping into three sections: CERN and its Member States, the
United States, and the remaining countries. Figures~11
and~12 present these results in graphical form, with
the contributions from CERN and its Member States  grouped.

In addition to the geographical distribution of the authors for
the major HEP journals, it is interesting to identify the most popular
journals of the single countries and institutions considered in this
analysis. To extract this information,  all articles with at
least one author from a given country or institution are first
selected. Then, the fraction of authorship of this country or
institution is calculated for each article. This fraction is assigned
to the journal where the article appeared. The sum of all these
fractions for each journal provides a score of the popularity of the
journal. If the sum of these scores is used to measure the total HEP
scientific production of the country, it can be used to normalise each
score and obtain the fractions of the HEP production of the country in
the different journals. The results of this study are presented in
Table~6 for each of the 22 countries and institutions
discussed in this article. The last three lines of Table~6
present the results summed over three groups: CERN and its Member
States, the United States and the remaining countries. The results for
these groups are presented in Figure~13. Figure~14 and~15 present results for some European
countries and institutions and Figure~16 presents results for
some of the remaining countries.

\section{Conclusions, with a Note on Open Access}

This article presents the results of the first
bibliometric study of HEP publishing which accounts for the
widespread phenomenon of coauthorship. The share of HEP scientific
results published by several countries and institutions is correctly
calculated and provides interesting insight into the collaborative
patterns within the HEP community. The publishing landscape of HEP is
further analysed to provide information on the journals most used by
the HEP community and on the geographical distribution of their authors.

It is interesting to put these results into the wider context of a
possible transition of HEP publishing to an Open 
Access model~\cite{taskForce}. The finding that 83\% of HEP articles are
published in just six journals and that 87\% of the articles appear in journals published by just
four publishers is particularly interesting. It 
demonstrates that the number of partners to be engaged with in
a debate on a change of the HEP publishing model is relatively small. 
The worldwide collaborative patterns in HEP,
which are quantified in this article, suggest that once a limited
number of countries embrace an Open Access publishing model, a
``domino effect'' likely to spread this policy to other countries, through
coauthorship links. Last, but not least, the  assessment of the
relative contribution to the worldwide production of HEP scientific
results which takes into account the coauthorship phenomenon,
presented in Table~2 and Figure~4, might constitute the
basis for a model where each country or institution would contribute
with their ``fair share'' towards the financial cost
of  Open Access publishing.

\section*{Acknowledgments}

The idea behind this analysis came up in many discussions with 
R\"udiger Voss and Gigi Rolandi on the topic of Open Access.
We are indebted to Sandrine Reyes and Susanne Sch\"afer for their help in the compilation
of the data set and to our colleagues at SLAC and elsewhere for maintaining and operating
{\tt SPIRES}.

\begin{sidewaystable}
  \begin{center}
    \begin{tabular}{|l|r|r|r|r|r|r|r|r|r|r|r|r|r|}
       \cline{2-14}
      \multicolumn{1}{c|}{}
      &\multicolumn{1}{c|}{\rotatebox{90}{Phys. Rev.}}&\multicolumn{1}{c|}{\rotatebox{90}{JHEP}}&\multicolumn{1}{c|}{\rotatebox{90}{Phys. Lett.}}&\multicolumn{1}{c|}{\rotatebox{90}{Nucl. Phys.}}&\multicolumn{1}{c|}{\rotatebox{90}{Phys. Rev.}\rotatebox{90}{Lett.}}&\multicolumn{1}{c|}{\rotatebox{90}{Eur. Phys J. }}&\multicolumn{1}{c|}{\rotatebox{90}{J. of Phys.}}&\multicolumn{1}{c|}{\rotatebox{90}{Mod. Phys.}\rotatebox{90}{Lett.}}&\multicolumn{1}{c|}{\rotatebox{90}{Int. J. Mod.}\rotatebox{90}{Phys.}}&\multicolumn{1}{c|}{\rotatebox{90}{Class. Quan.}\rotatebox{90}{Grav.}}&\multicolumn{1}{c|}{\rotatebox{90}{JCAP}}&\multicolumn{1}{c|}{\rotatebox{90}{NIM}}&\multicolumn{1}{c|}{\rotatebox{90}{Others}}\\
      \hline
      CERN&0.7\%&2.0\%&1.5\%&2.5\%&0.7\%&2.4\%&1.3\%&0.4\%&0.3\%&1.4\%&2.0\%&0.3\%&1.1\%\\
      Germany&7.2\%&9.3\%&9.3\%&13.5\%&9.0\%&14.1\%&6.1\%&4.2\%&5.4\%&6.5\%&7.8\%&1.6\%&8.5\%\\
      UK&6.1\%&10.4\%&6.5\%&7.8\%&5.1\%&9.8\%&10.6\%&3.5\%&5.1\%&16.6\%&12.6\%&16.0\%&4.4\%\\
      INFN&4.5\%&6.9\%&5.8\%&10.3\%&5.4\%&5.4\%&4.6\%&4.0\%&5.3\%&4.3\%&4.4\%&14.6\%&5.7\%\\
      France&2.5\%&2.3\%&4.0\%&5.3\%&2.8\%&5.4\%&3.4\%&1.7\%&2.2\%&1.0\%&4.8\%&3.0\%&5.0\%\\
      Spain&2.6\%&4.8\%&2.3\%&2.2\%&2.9\%&1.8\%&2.3\%&0.9\%&1.0\%&1.8\%&5.3\%&0.2\%&1.8\%\\
      Switzerland&0.6\%&1.3\%&1.9\%&1.8\%&0.9\%&0.5\%&2.2\%&$-$&$-$&$-$&2.3\%&0.2\%&0.5\%\\
      Sweden&0.6\%&1.6\%&1.0\%&1.2\%&0.7\%&1.0\%&$-$&0.4\%&$-$&$-$&4.9\%&$-$&0.6\%\\
      Portugal&1.1\%&0.5\%&1.0\%&0.6\%&1.2\%&0.2\%&1.7\%&1.2\%&$-$&2.3\%&$-$&$-$&1.0\%\\
      Netherlands&0.4\%&2.2\%&0.5\%&1.1\%&0.4\%&0.4\%&$-$&$-$&1.1\%&1.6\%&2.9\%&0.2\%&0.7\%\\
      Other M.S.&5.6\%&7.9\%&6.1\%&7.6\%&4.0\%&10.4\%&9.6\%&3.5\%&9.3\%&9.0\%&9.0\%&10.9\%&8.1\%\\
      Russia&3.9\%&1.5\%&5.7\%&3.9\%&2.0\%&8.6\%&5.7\%&7.4\%&4.9\%&2.3\%&0.8\%&12.6\%&15.2\%\\
      Israel&1.2\%&1.3\%&0.7\%&1.4\%&0.3\%&1.1\%&$-$&1.2\%&1.1\%&1.5\%&0.7\%&0.2\%&0.1\%\\
      United States&30.8\%&24.3\%&19.2\%&21.0\%&48.1\%&6.9\%&10.8\%&16.8\%&23.0\%&24.4\%&16.3\%&31.7\%&10.8\%\\
      Canada&3.0\%&3.0\%&2.0\%&3.6\%&2.8\%&0.7\%&3.9\%&3.4\%&0.3\%&7.1\%&2.6\%&1.4\%&1.0\%\\
      Brazil&2.8\%&1.7\%&3.3\%&0.2\%&0.7\%&5.3\%&6.4\%&5.4\%&5.1\%&1.1\%&1.8\%&1.0\%&3.0\%\\
      Japan&8.3\%&5.8\%&7.9\%&7.2\%&4.9\%&2.4\%&3.1\%&4.3\%&9.4\%&4.2\%&11.3\%&1.6\%&13.6\%\\
      China&5.6\%&1.9\%&5.8\%&1.8\%&2.2\%&10.7\%&7.5\%&4.6\%&3.0\%&2.3\%&4.1\%&$-$&6.8\%\\
      India&2.4\%&2.4\%&3.6\%&1.5\%&1.1\%&2.8\%&5.2\%&7.0\%&6.3\%&2.7\%&5.4\%&$-$&1.7\%\\
      Taiwan&1.8\%&0.5\%&1.5\%&0.6\%&1.4\%&1.6\%&$-$&1.7\%&$-$&$-$&$-$&$-$&0.8\%\\
      Korea&1.8\%&2.6\%&2.6\%&0.9\%&1.3\%&0.5\%&$-$&3.6\%&$-$&1.1\%&$-$&0.7\%&0.8\%\\
      Other Countries&6.5\%&5.8\%&7.8\%&4.1\%&2.3\%&7.9\%&15.7\%&24.6\%&17.0\%&8.6\%&1.0\%&3.8\%&8.9\%\\
      \hline
      Total & 100\%&100\%&100\%&100\%&100\%&100\%&100\%&100\%&100\%&100\%&100\%&100\%&100\%\\
      \hline
      \hline
      CERN \& &&&&&&&&&&&&&\\
      Member States&31.9\%&49.2\%&40.1\%&53.8\%&33.0\%&51.5\%&41.7\%&19.9\%&29.7\%&44.5\%&56.0\%&47.0\%&37.3\%\\
      United States&30.8\%&24.3\%&19.2\%&21.0\%&48.1\%&6.9\%&10.8\%&16.8\%&23.0\%&24.4\%&16.3\%&31.7\%&10.8\%\\
      Other Countries&37.3\%&26.4\%&40.7\%&25.2\%&18.9\%&41.6\%&47.5\%&63.3\%&47.3\%&31.0\%&27.6\%&21.3\%&51.9\%\\
      \hline
      Total & 100\%&100\%&100\%&100\%&100\%&100\%&100\%&100\%&100\%&100\%&100\%&100\%&100\%\\
      \hline
    \end{tabular}
    \caption{Geographical distribution of the authors of HEP
      journals. The lower part of the table summarises the results for
      three sections of the HEP community:
      CERN and its Member States, the United States, and the 
      remaining countries.}
  \end{center}
  \label{tab:5}
\end{sidewaystable}

\begin{sidewaystable}
  \begin{center}
    \begin{tabular}{|lrrrrrrrrrrrrr|r|}
      \cline{2-15}
      \multicolumn{1}{c|}{}
      &\multicolumn{1}{c|}{\rotatebox{90}{Phys. Rev.}}&\multicolumn{1}{c|}{\rotatebox{90}{JHEP}}&\multicolumn{1}{c|}{\rotatebox{90}{Phys. Lett.}}&\multicolumn{1}{c|}{\rotatebox{90}{Nucl. Phys.}}&\multicolumn{1}{c|}{\rotatebox{90}{Phys. Rev.}\rotatebox{90}{Lett.}}&\multicolumn{1}{c|}{\rotatebox{90}{Eur. Phys J. }}&\multicolumn{1}{c|}{\rotatebox{90}{J. of Phys.}}&\multicolumn{1}{c|}{\rotatebox{90}{Mod. Phys.}\rotatebox{90}{Lett.}}&\multicolumn{1}{c|}{\rotatebox{90}{Int. J. Mod.}\rotatebox{90}{Phys.}}&\multicolumn{1}{c|}{\rotatebox{90}{Class. Quan.}\rotatebox{90}{Grav.}}&\multicolumn{1}{c|}{\rotatebox{90}{JCAP}}&\multicolumn{1}{c|}{\rotatebox{90}{NIM}}&\multicolumn{1}{c|}{\rotatebox{90}{Others}}&\multicolumn{1}{c|}{\rotatebox{90}{Total}}\\
      \hline
      CERN&15.5\%&29.3\%&16.4\%&15.6\%&2.4\%&7.7\%&2.2\%&0.7\%&0.4\%&1.9\%&1.5\%&$-$&6.3\%&100\%\\
      \hline
      Germany&26.1\%&20.4\%&15.4\%&12.8\%&4.9\%&6.9\%&1.6\%&1.1\%&1.1\%&1.3\%&0.9\%&$-$&7.5\%&100\%\\
      \hline
      UK&26.2\%&27.0\%&12.8\%&8.8\%&3.3\%&5.7\%&3.3\%&1.1\%&1.2\%&3.9\%&1.7\%&0.5\%&4.5\%&100\%\\
      \hline
      INFN&24.4\%&22.7\%&14.6\%&14.7\%&4.4\%&4.0\%&1.8\%&1.6\%&1.6\%&1.3\%&0.8\%&0.5\%&7.5\%&100\%\\
      \hline
      France&24.7\%&13.4\%&18.2\%&13.5\%&4.1\%&7.2\%&2.4\%&1.2\%&1.2\%&0.5\%&1.5\%&0.2\%&11.8\%&100\%\\
      \hline
      Spain&29.1\%&33.3\%&12.2\%&6.5\%&5.0\%&2.8\%&1.9\%&0.7\%&0.7\%&1.1\%&1.9\%&$-$&4.8\%&100\%\\
      \hline
      Switzerland&18.3\%&24.0\%&26.4\%&14.4\%&4.1\%&2.1\%&4.7\%&$-$&$-$&$-$&2.2\%&$-$&3.8\%&100\%\\
      \hline
      Sweden&19.8\%&33.5\%&15.7\%&10.9\%&3.6\%&4.9\%&$-$&1.1\%&$-$&$-$&5.5\%&$-$&5.0\%&100\%\\
      \hline
      Portugal&39.0\%&10.7\%&16.8\%&5.7\%&6.5\%&1.1\%&4.4\%&3.1\%&$-$&4.4\%&$-$&$-$&8.3\%&100\%\\
      \hline
      Netherlands&14.8\%&47.0\%&8.7\%&10.6\%&2.2\%&1.9\%&$-$&$-$&2.2\%&3.1\%&3.4\%&0.1\%&6.0\%&100\%\\
      \hline
      Other M.S.&26.2\%&22.2\%&13.0\%&9.3\%&2.8\%&6.6\%&3.2\%&1.2\%&2.4\%&2.3\%&1.3\%&0.4\%&9.1\%&100\%\\
      \hline
      Russia&25.8\%&6.1\%&17.3\%&6.8\%&2.0\%&7.8\%&2.8\%&3.6\%&1.8\%&0.8\%&0.2\%&0.6\%&24.4\%&100\%\\
      \hline
      Israel&38.5\%&25.2\%&10.2\%&11.4\%&1.4\%&4.8\%&$-$&2.6\%&1.9\%&2.6\%&0.6\%&0.1\%&0.7\%&100\%\\
      \hline
      United States&40.5\%&19.4\%&11.6\%&7.3\%&9.6\%&1.2\%&1.0\%&1.6\%&1.7\%&1.8\%&0.7\%&0.3\%&3.4\%&100\%\\
      \hline
      Canada&35.2\%&21.3\%&10.9\%&11.2\%&5.1\%&1.1\%&3.4\%&2.9\%&0.2\%&4.6\%&1.0\%&0.1\%&3.0\%&100\%\\
      \hline
      Brazil&34.4\%&12.4\%&18.3\%&0.7\%&1.3\%&8.8\%&5.7\%&4.8\%&3.4\%&0.8\%&0.7\%&0.1\%&8.8\%&100\%\\
      \hline
      Japan&35.6\%&15.0\%&15.5\%&8.1\%&3.2\%&1.4\%&1.0\%&1.3\%&2.2\%&1.0\%&1.6\%&$-$&14.1\%&100\%\\
      \hline
      China&38.4\%&7.9\%&18.1\%&3.3\%&2.3\%&10.0\%&3.7\%&2.3\%&1.1\%&0.9\%&0.9\%&$-$&11.2\%&100\%\\
      \hline
      India&28.3\%&17.2\%&19.8\%&4.6\%&1.9\%&4.6\%&4.5\%&6.0\%&4.2\%&1.8\%&2.1\%&$-$&4.9\%&100\%\\
      \hline
      Taiwan&48.0\%&8.8\%&18.3\%&4.6\%&5.6\%&6.1\%&$-$&3.4\%&$-$&$-$&$-$&$-$&5.1\%&100\%\\
      \hline
      Korea&32.7\%&27.8\%&21.0\%&4.4\%&3.6\%&1.2\%&$-$&4.7\%&$-$&1.1\%&$-$&0.1\%&3.5\%&100\%\\
      \hline
      Other Countries&28.7\%&15.5\%&15.9\%&4.7\%&1.5\%&4.7\%&5.1\%&7.9\%&4.2\%&2.1\%&0.1\%&0.1\%&9.6\%&100\%\\
      \hline
      \hline
      CERN \&  &&&&&&&&&&&&&&\\
      Member States&25.3\%&23.7\%&14.6\%&11.2\%&4.0\%&5.6\%&2.4\%&1.1\%&1.3\%&1.9\%&1.4\%&0.3\%&7.2\%&100\%\\
      \hline
      United States&40.5\%&19.4\%&11.6\%&7.3\%&9.6\%&1.2\%&1.0\%&1.6\%&1.7\%&1.8\%&0.7\%&0.3\%&3.4\%&100\%\\
      \hline
      Other Countries&33.0\%&14.2\%&16.5\%&5.9\%&2.5\%&5.0\%&3.1\%&4.0\%&2.3\%&1.5\%&0.8\%&0.1\%&11.1\%&100\%\\
      \hline
    \end{tabular}
    \caption{Distribution of each country's HEP articles 
      in several journals.  The lower part of the table summarises the
      results for
      CERN and its Member States, the United States, and all remaining
      countries.}
  \end{center}
  \label{tab:6}
\end{sidewaystable}

\clearpage

\begin{figure}[p]
  \begin{center}
    \begin{tabular}{cc}
      \mbox{\includegraphics[width=0.5\textwidth]{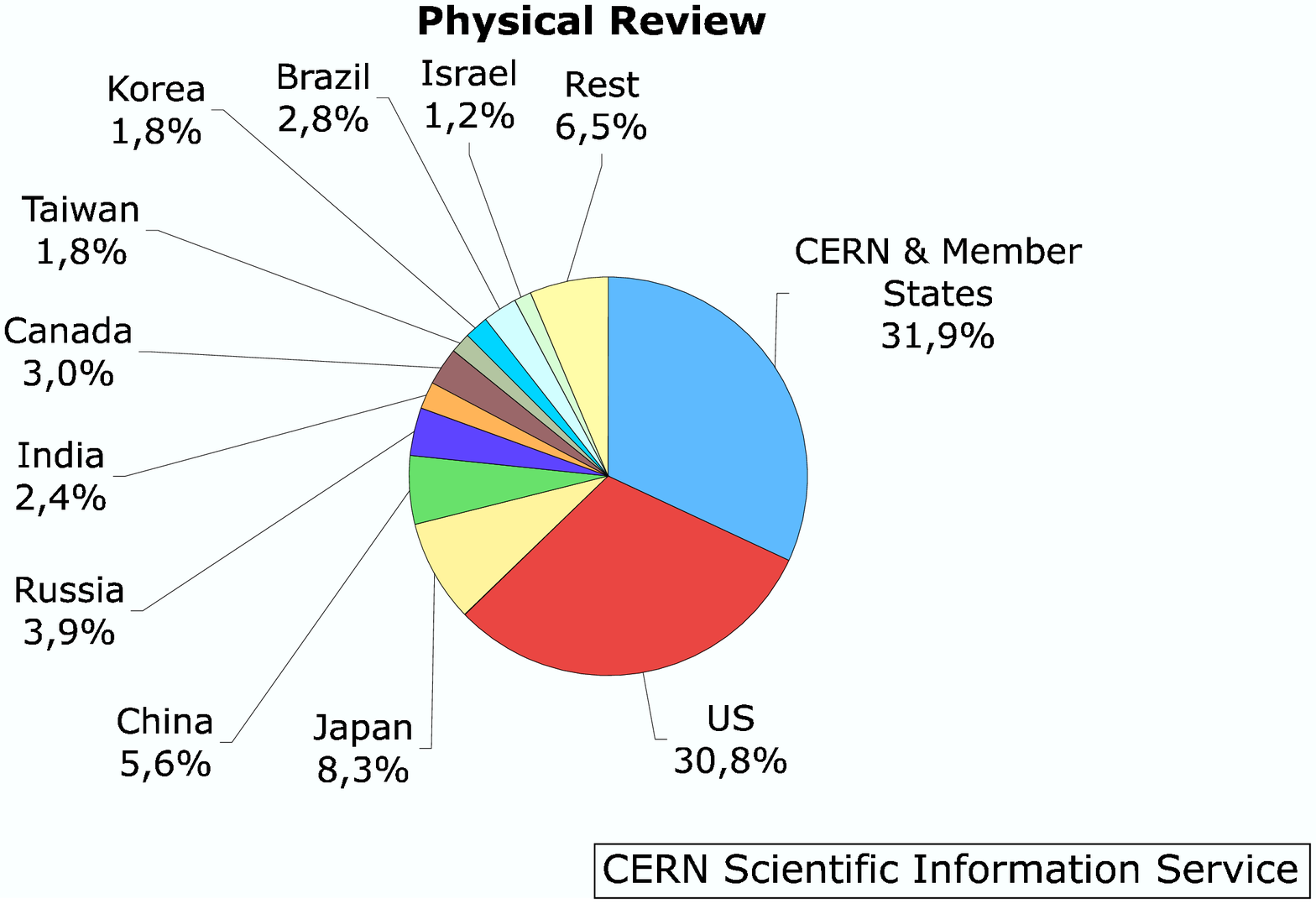}}&
      \mbox{\includegraphics[width=0.5\textwidth]{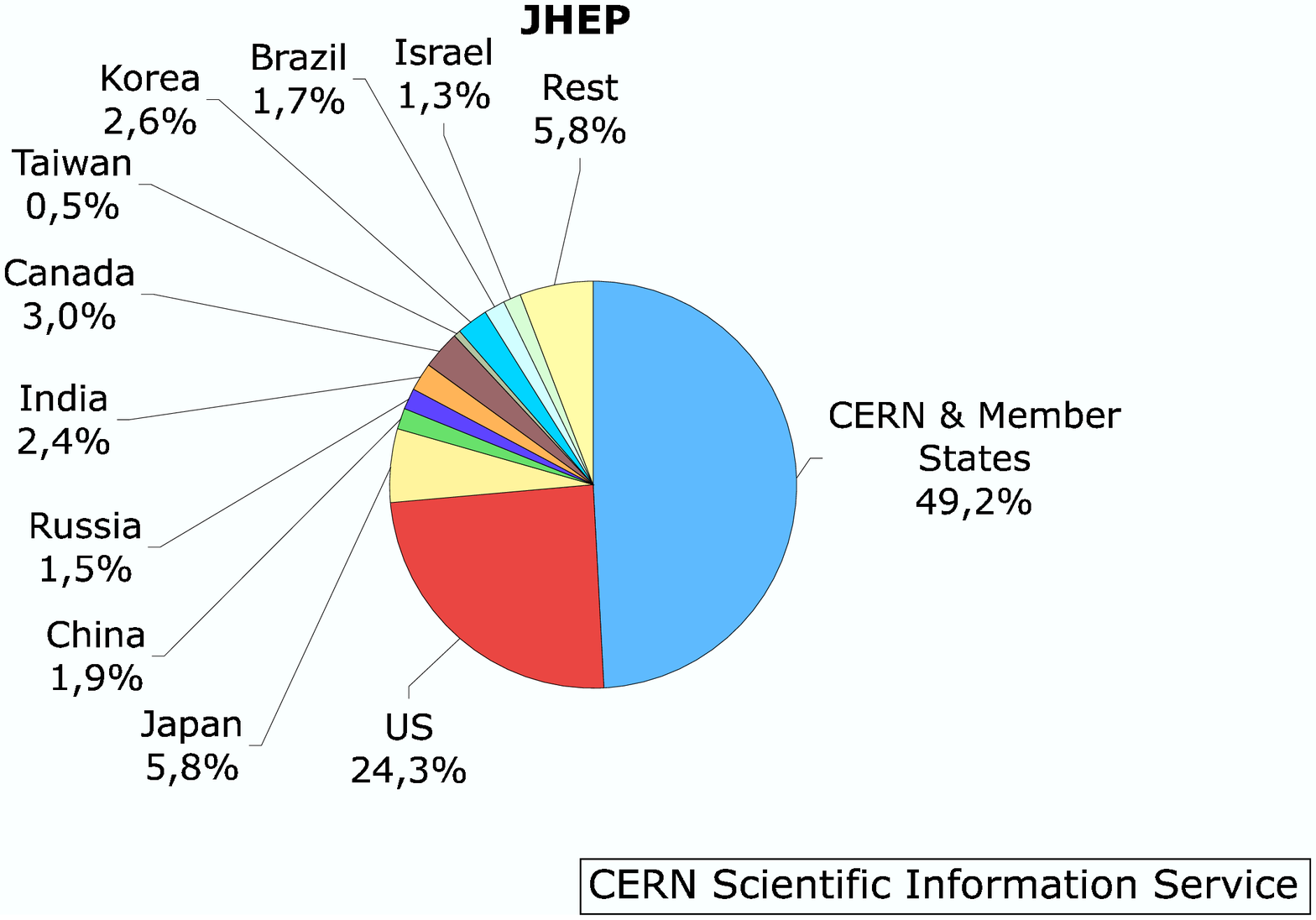}}\\
      \mbox{\includegraphics[width=0.5\textwidth]{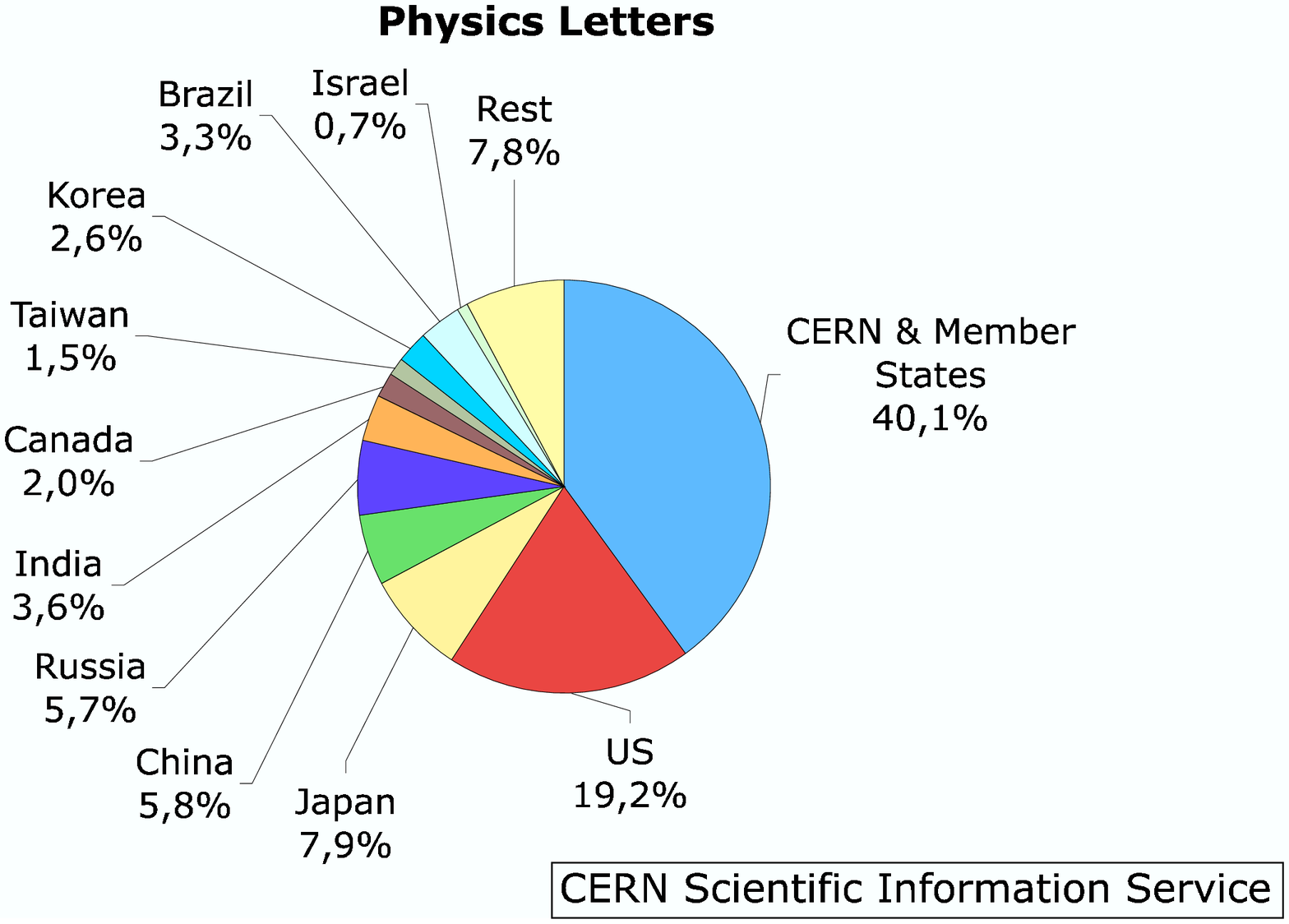}}&
      \mbox{\includegraphics[width=0.5\textwidth]{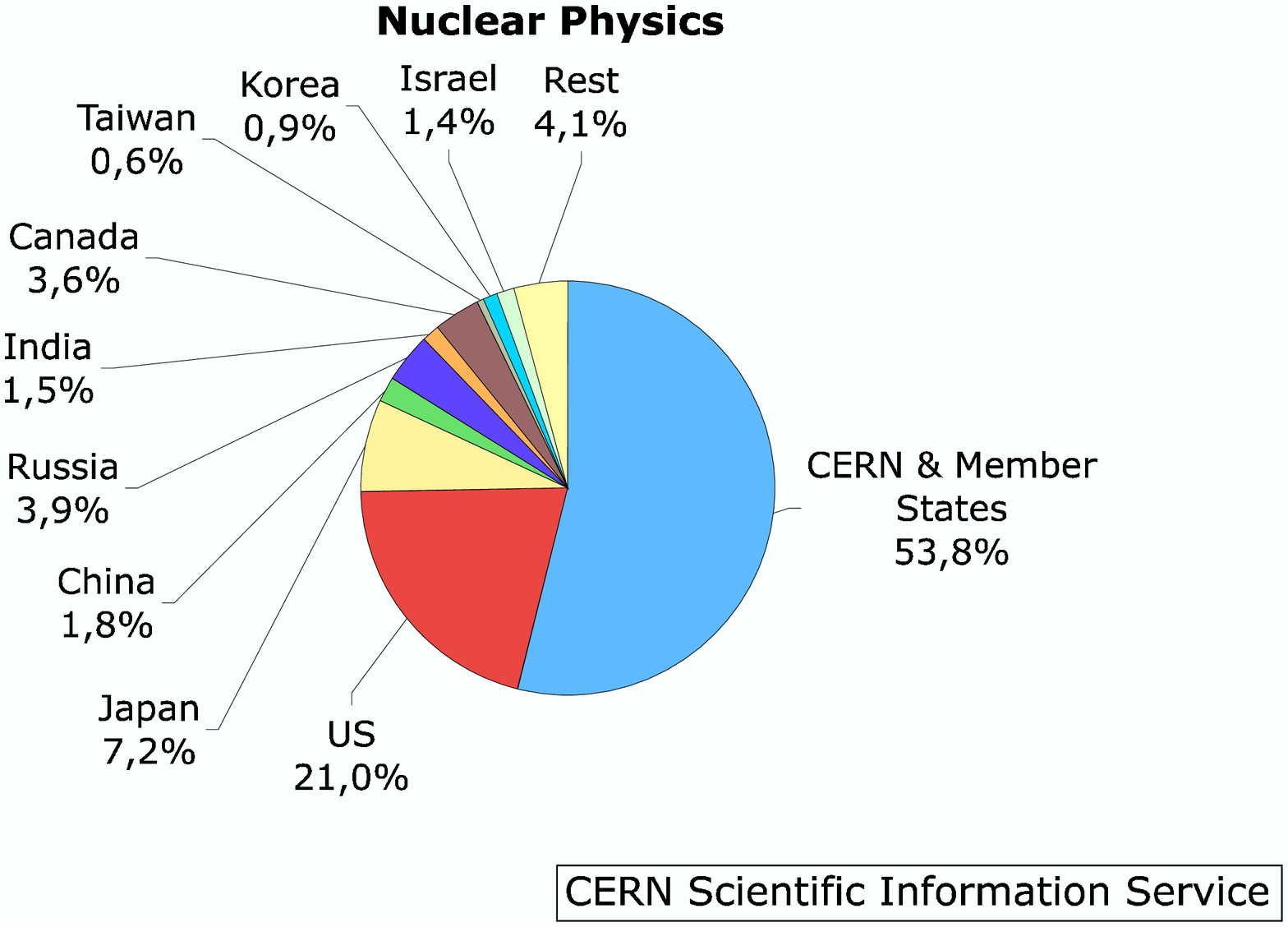}}\\
      \mbox{\includegraphics[width=0.5\textwidth]{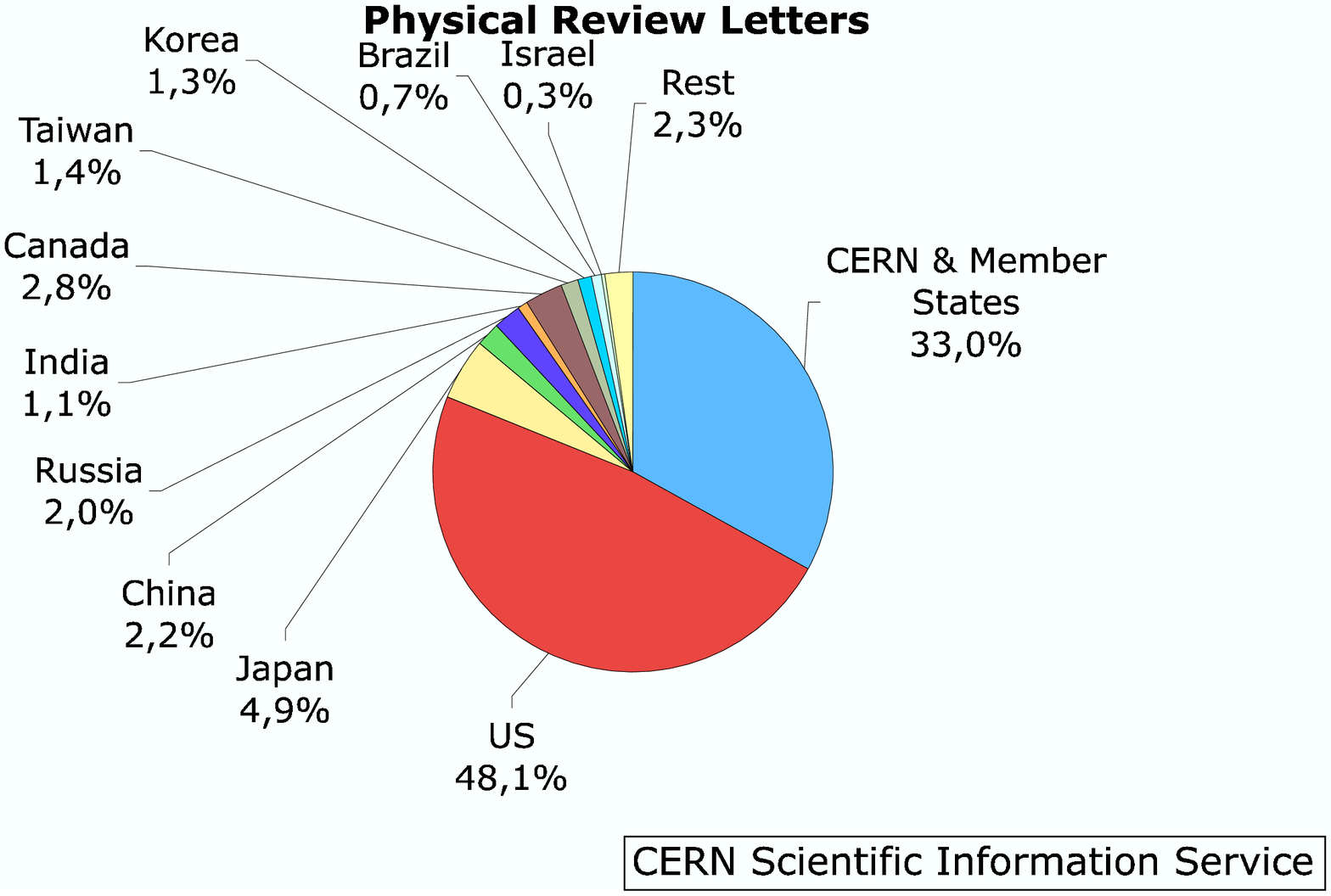}}&
      \mbox{\includegraphics[width=0.5\textwidth]{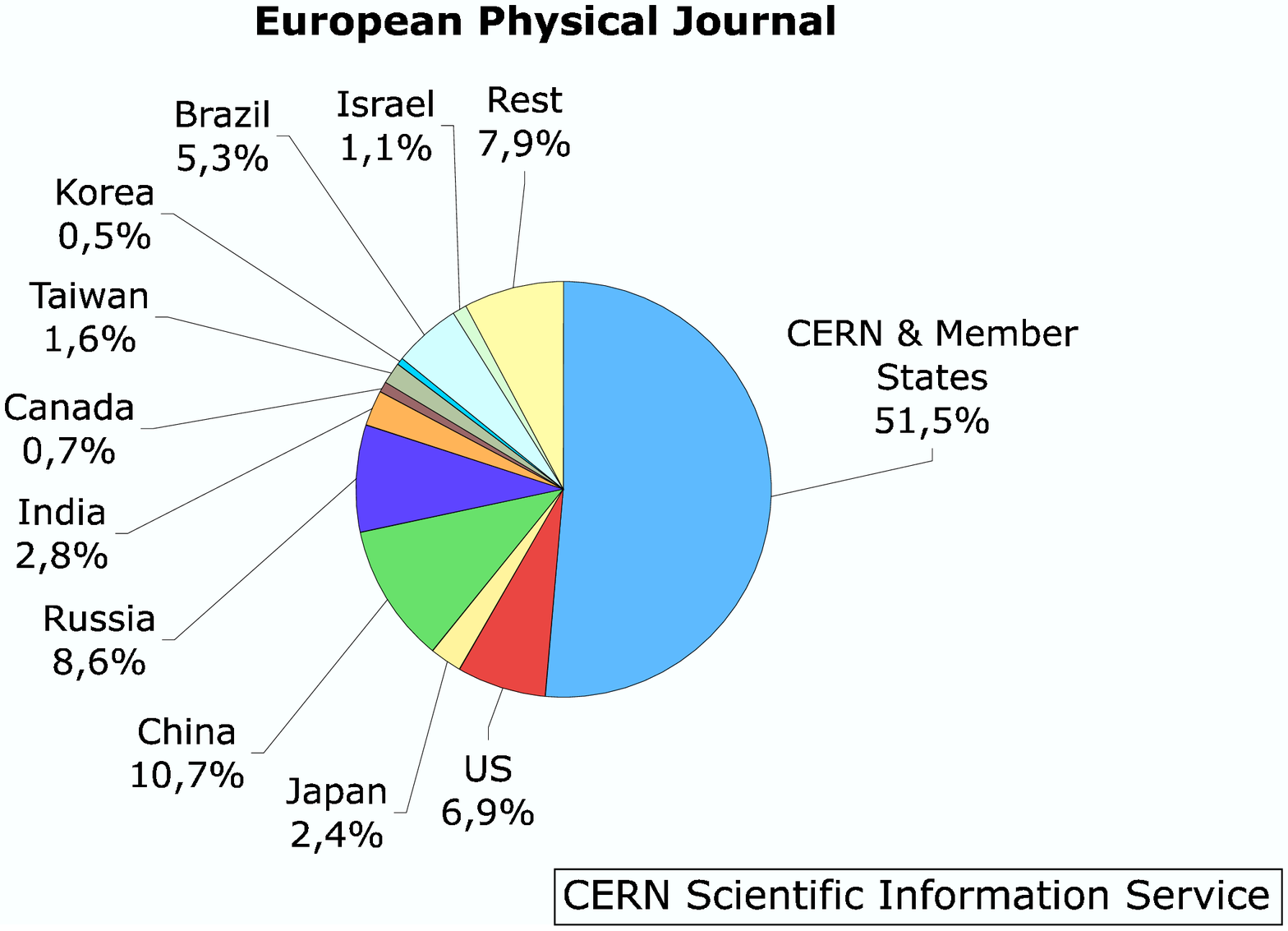}}\\
      \end{tabular}
    \caption{Geographical distribution of HEP authors by journals.}
  \end{center}
  \label{fig:6}
\end{figure}\clearpage

\begin{figure}[p]
  \begin{center}
    \begin{tabular}{cc}
      \mbox{\includegraphics[width=0.5\textwidth]{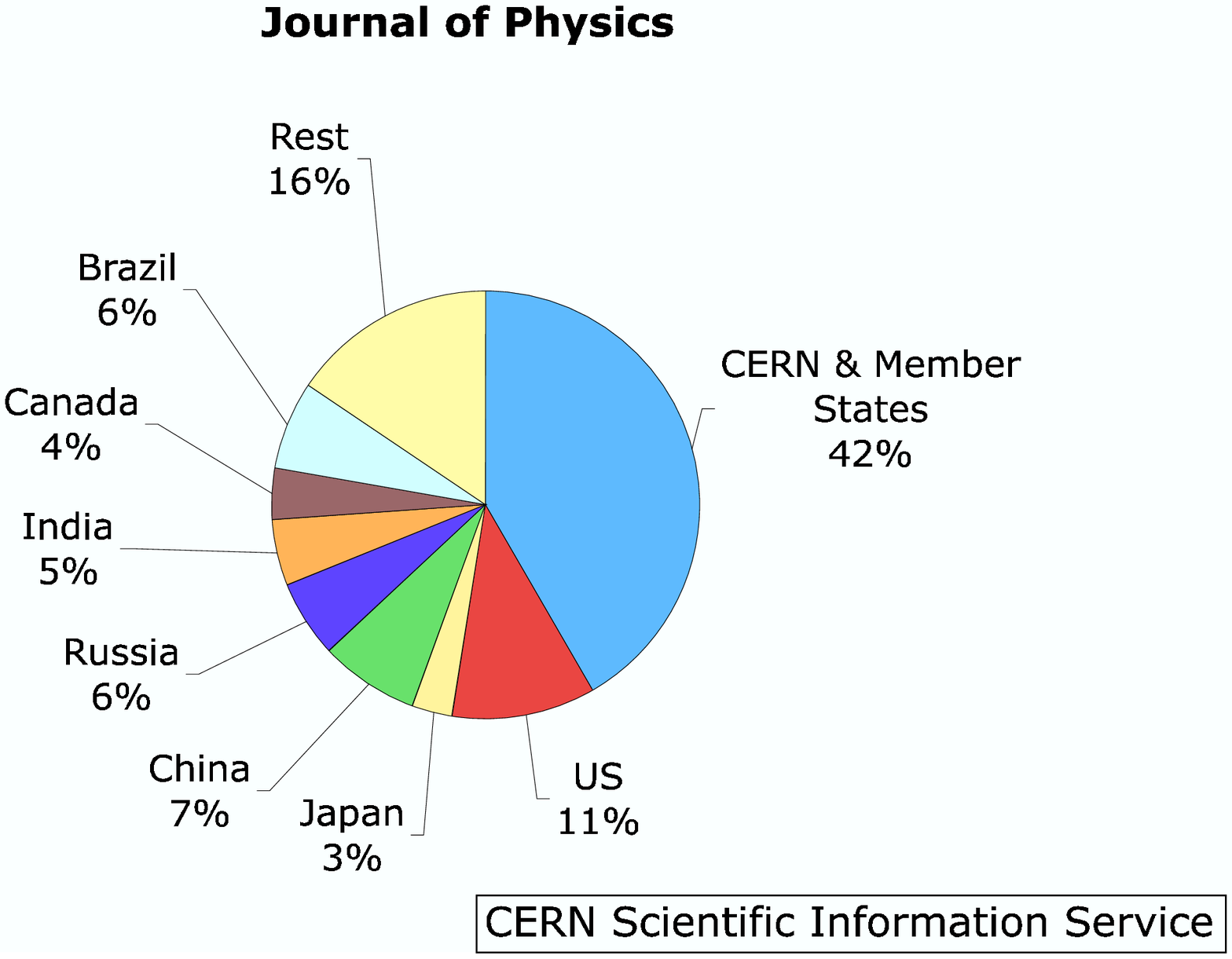}}&
      \mbox{\includegraphics[width=0.5\textwidth]{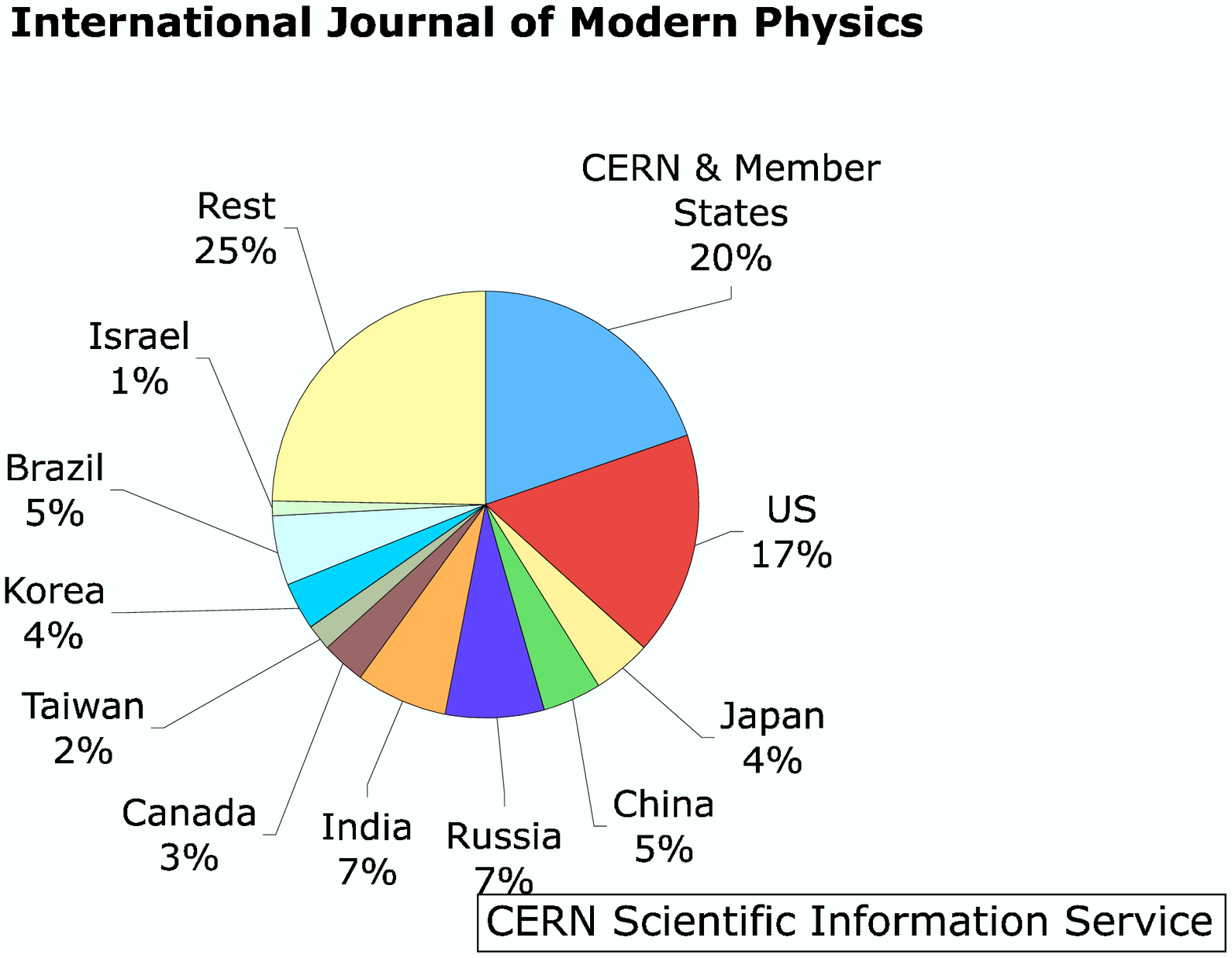}}\\
      \mbox{\includegraphics[width=0.5\textwidth]{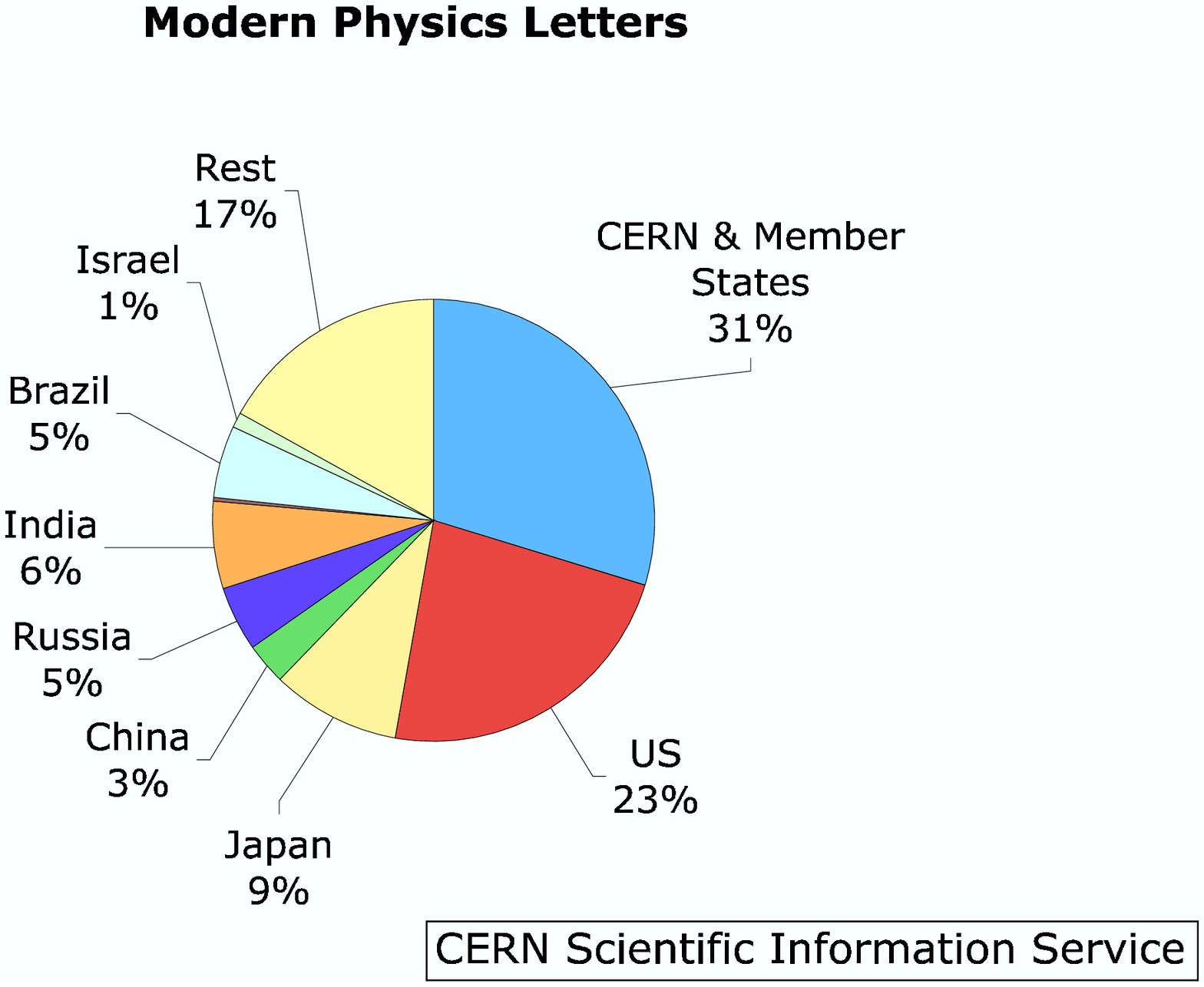}}&
      \mbox{\includegraphics[width=0.5\textwidth]{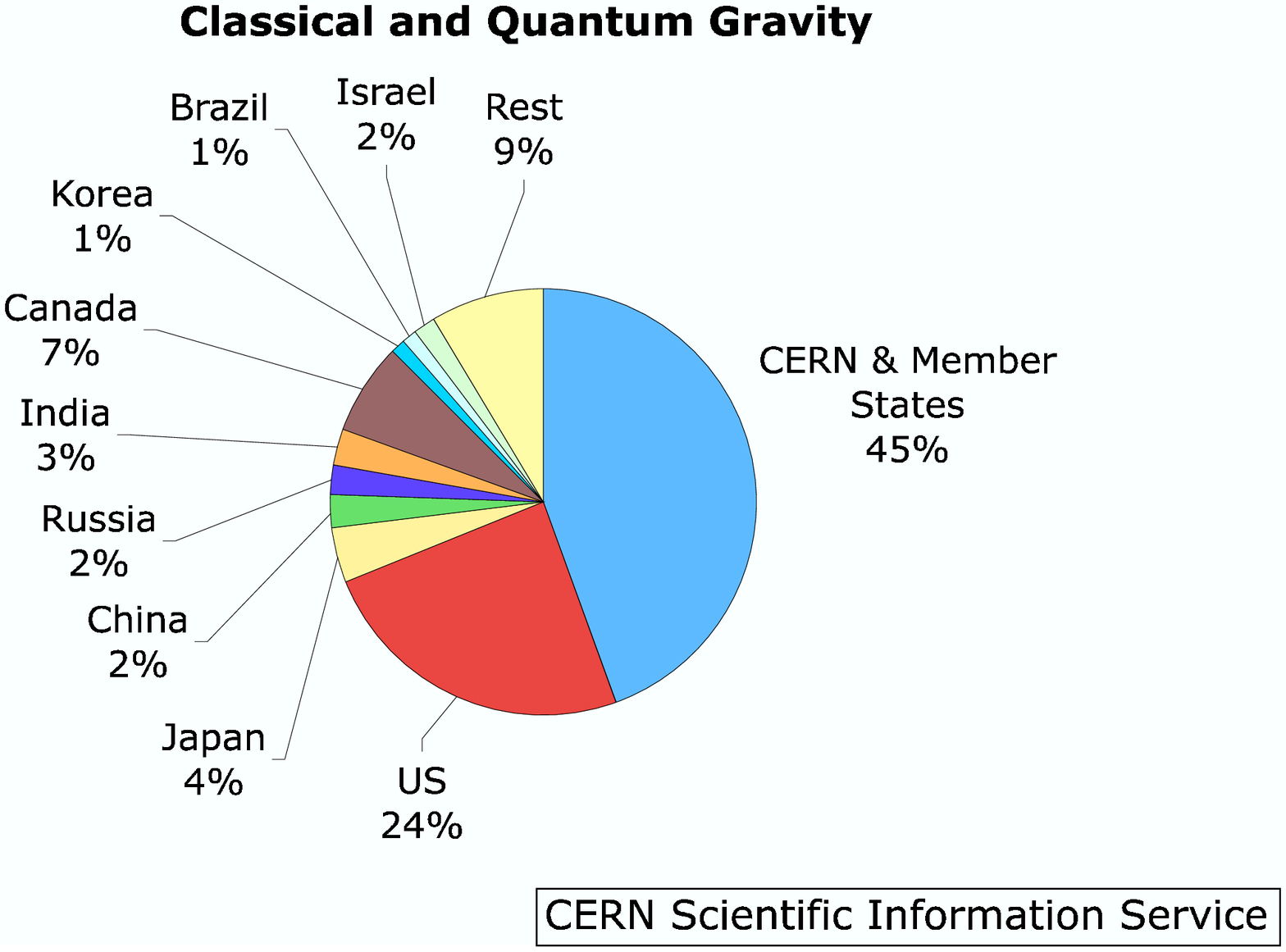}}\\
      \mbox{\includegraphics[width=0.5\textwidth]{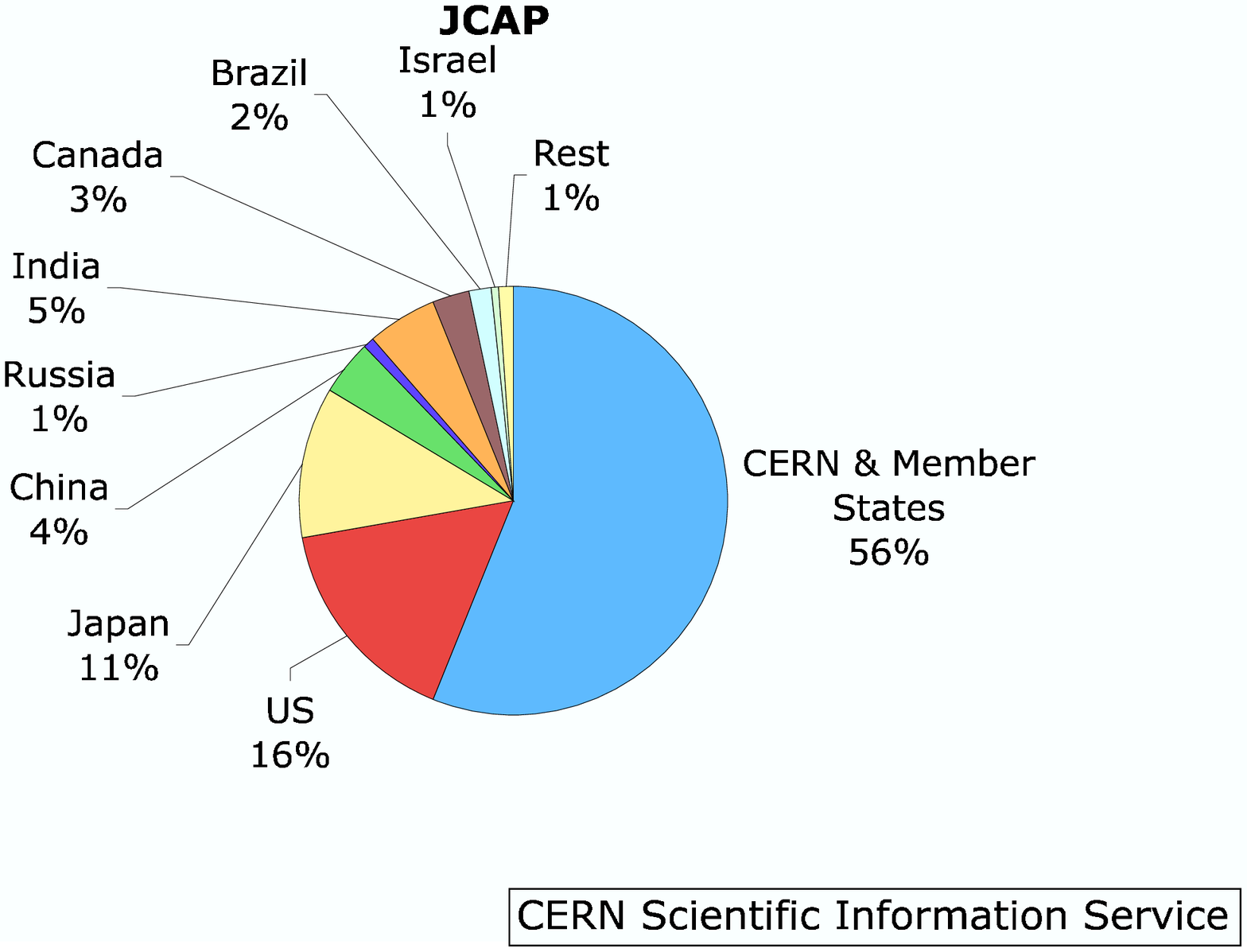}}&
      \mbox{\includegraphics[width=0.5\textwidth]{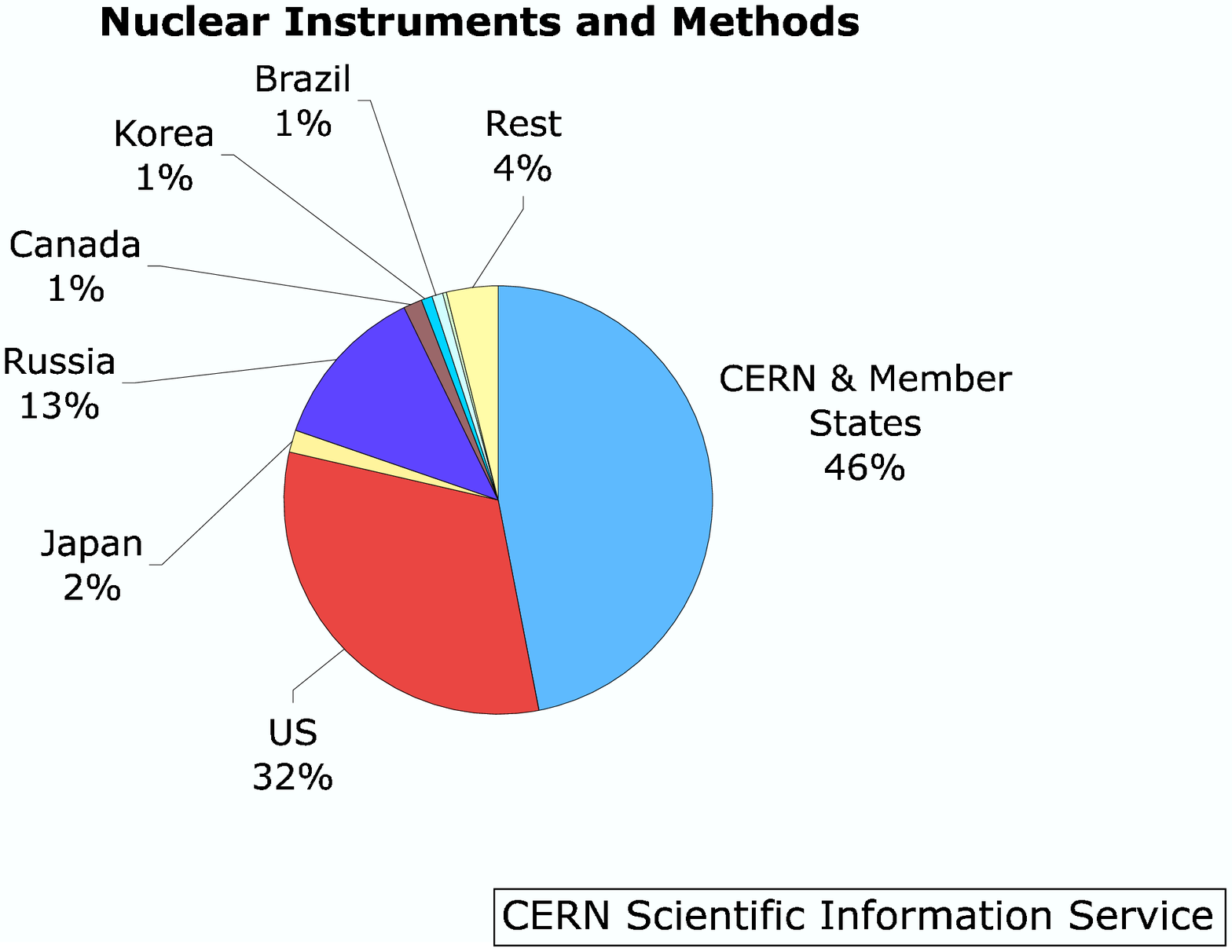}}\\
      \end{tabular}
    \caption{Geographical distribution of HEP authors by journals.}
  \end{center}
  \label{fig:7}
\end{figure}\clearpage

\begin{figure}[htb]
  \begin{center}
    \begin{tabular}{c}
      \mbox{\includegraphics[width=0.6\textwidth]{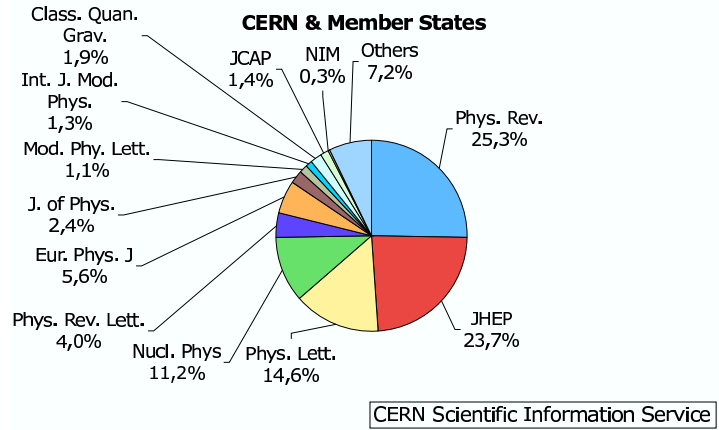}}\\
      \mbox{\includegraphics[width=0.6\textwidth]{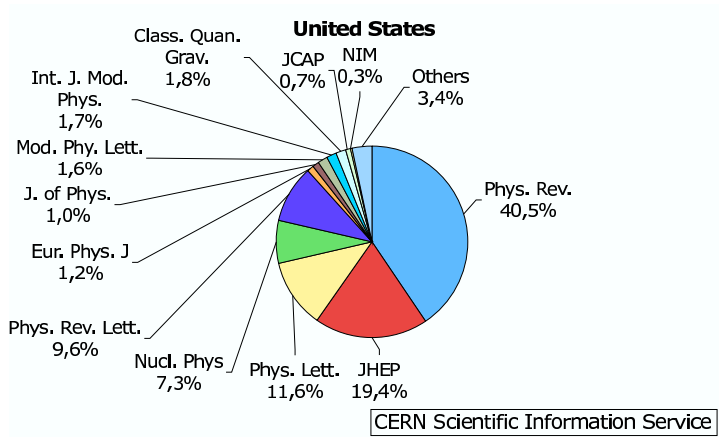}}\\
      \mbox{\includegraphics[width=0.6\textwidth]{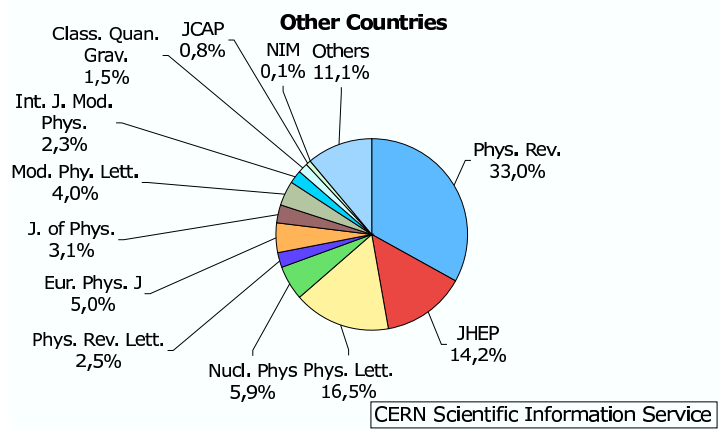}}\\
      \end{tabular}
    \caption{Distribution of HEP articles in different journals
      for three country groups.}
  \end{center}
  \label{fig:8}
\end{figure}\clearpage

\begin{figure}[p]
  \begin{center}
    \begin{tabular}{cc}
      \mbox{\includegraphics[width=0.5\textwidth]{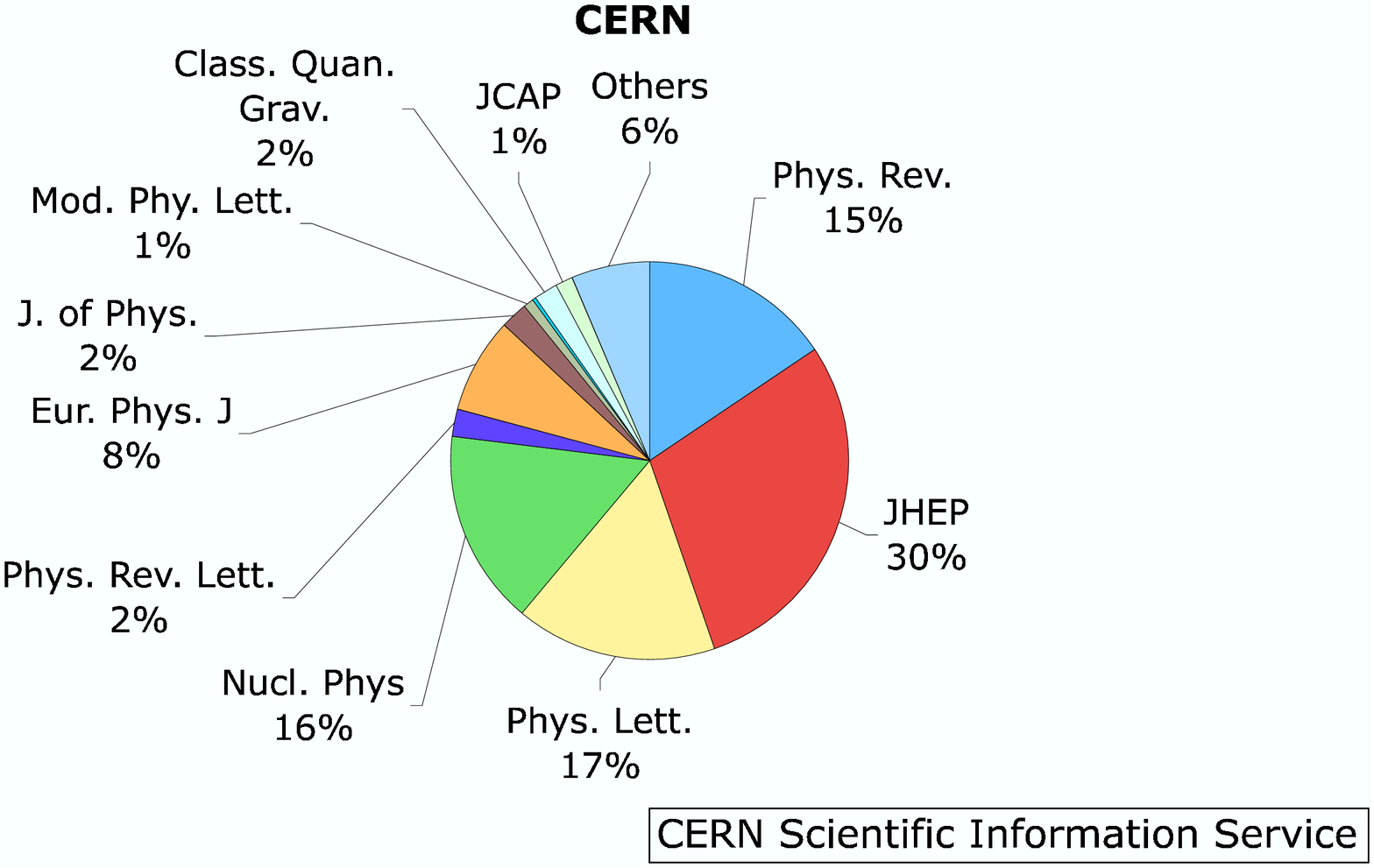}}&
      \mbox{\includegraphics[width=0.5\textwidth]{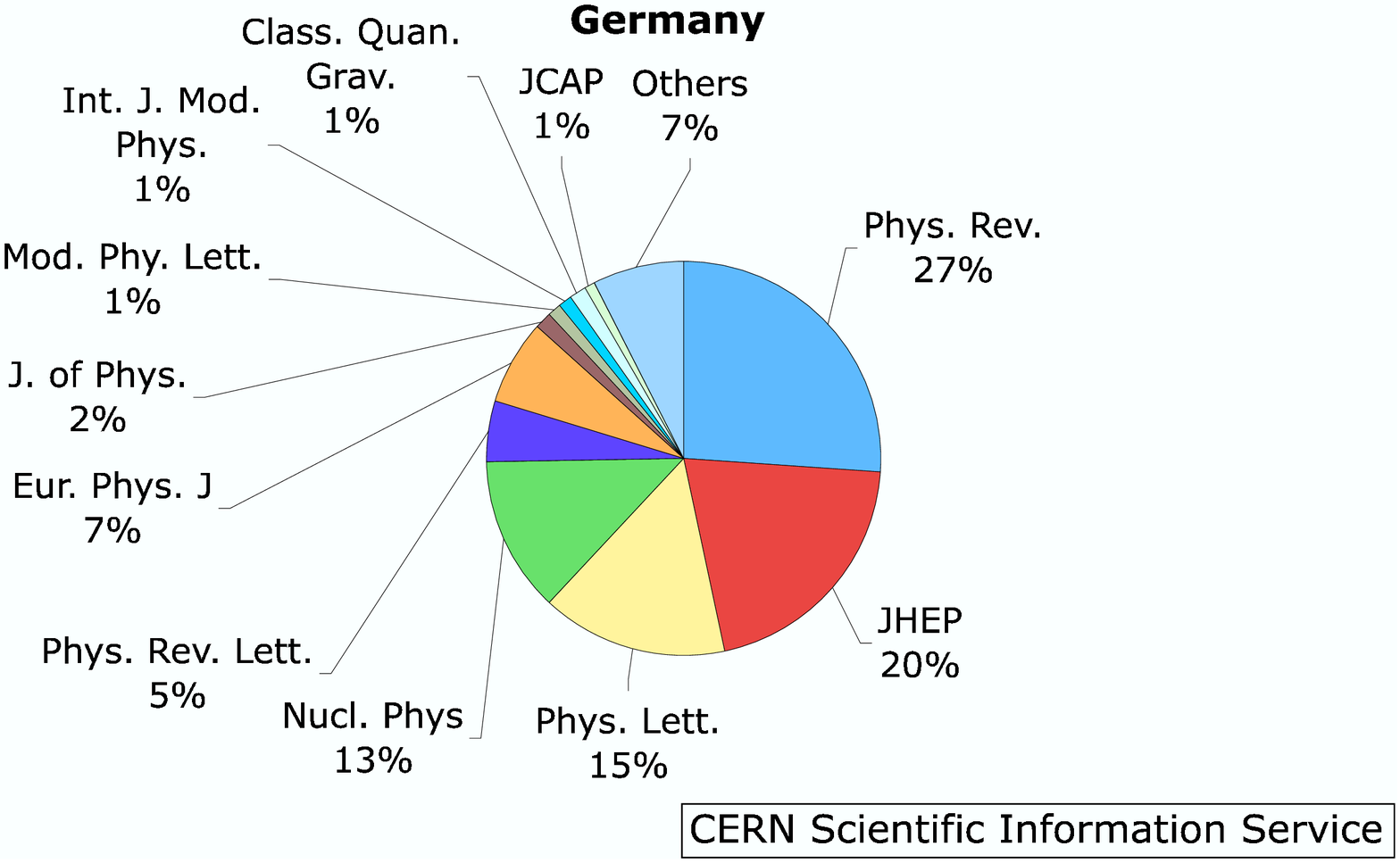}}\\
      \mbox{\includegraphics[width=0.5\textwidth]{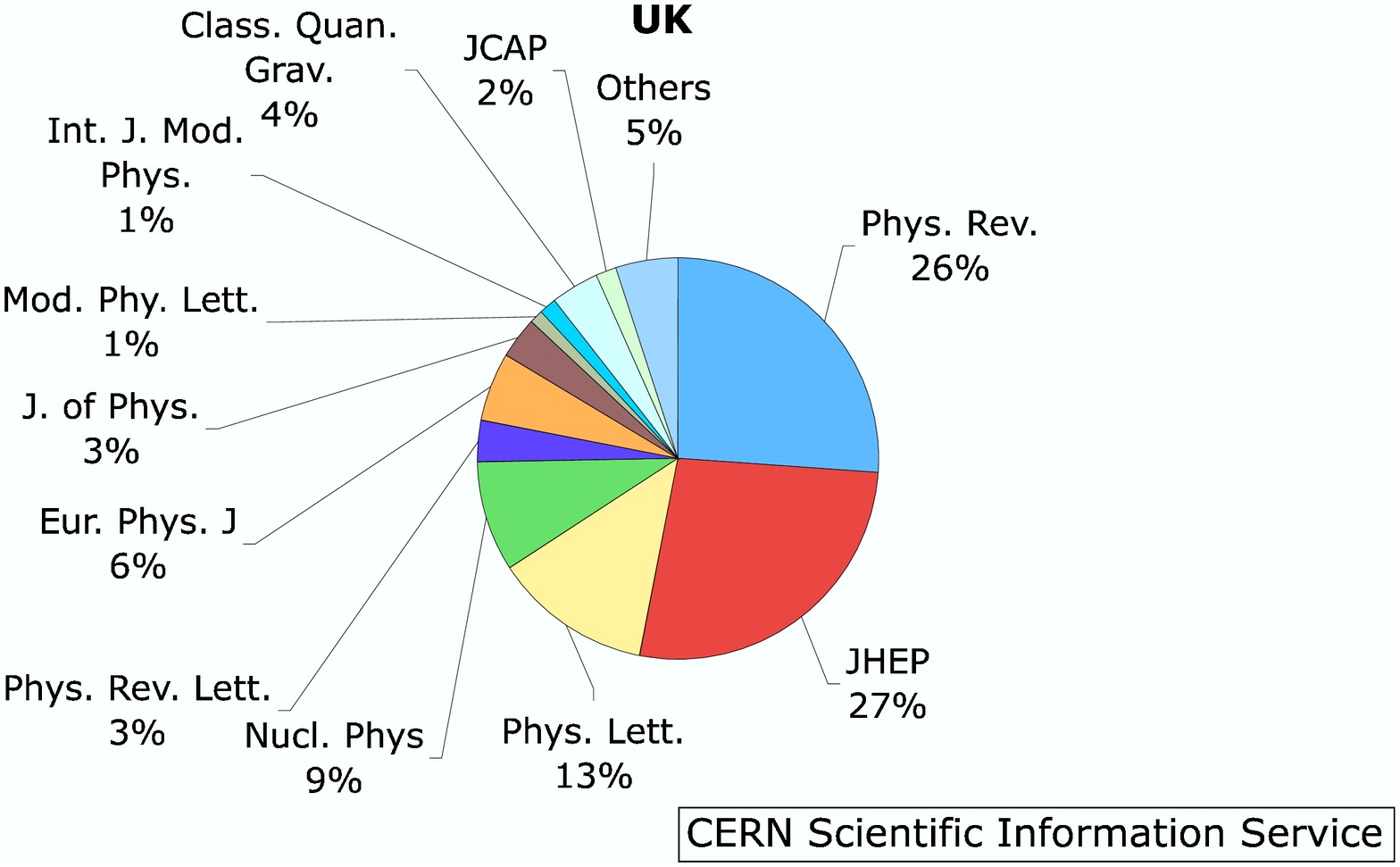}}&
      \mbox{\includegraphics[width=0.5\textwidth]{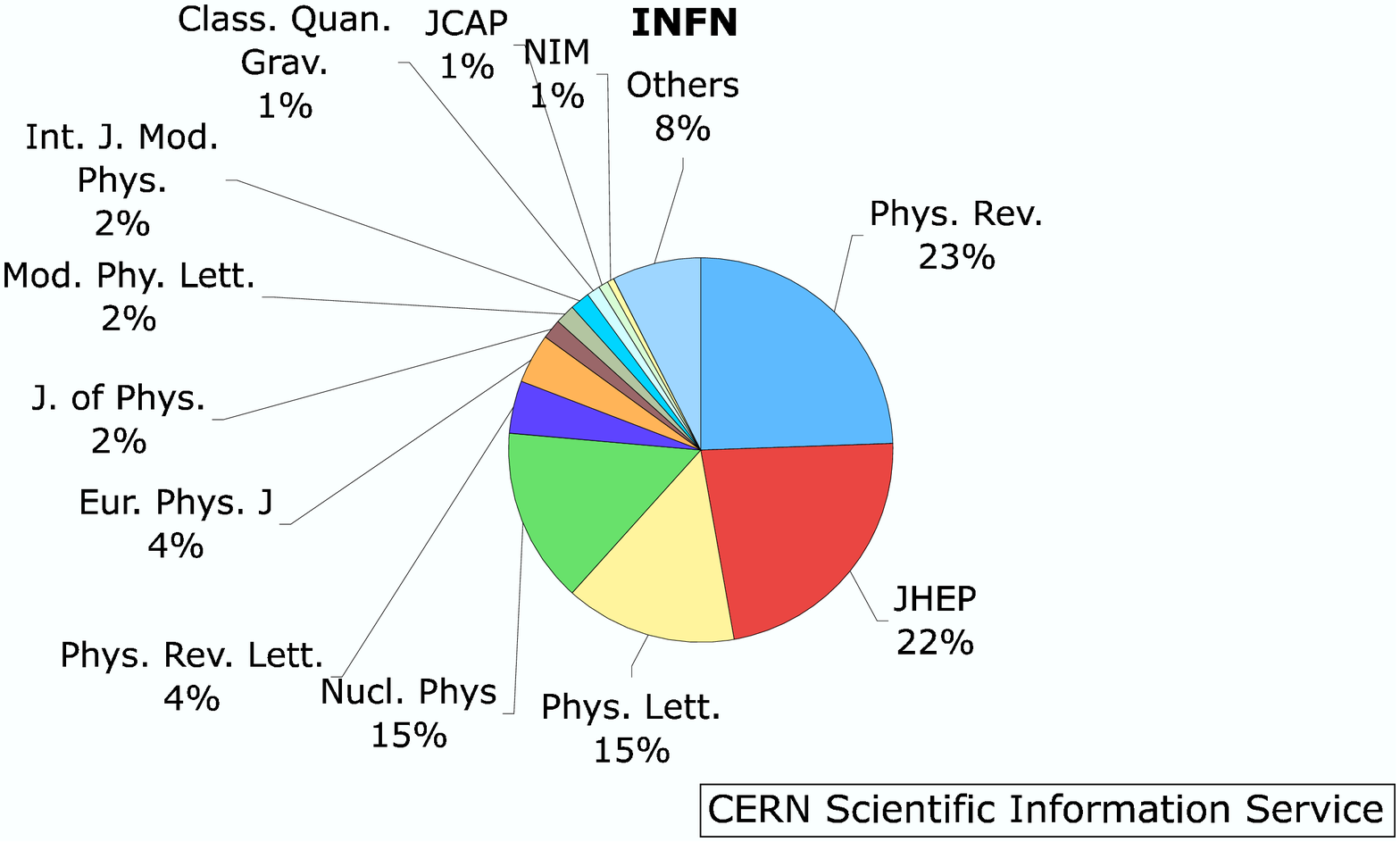}}\\
      \mbox{\includegraphics[width=0.5\textwidth]{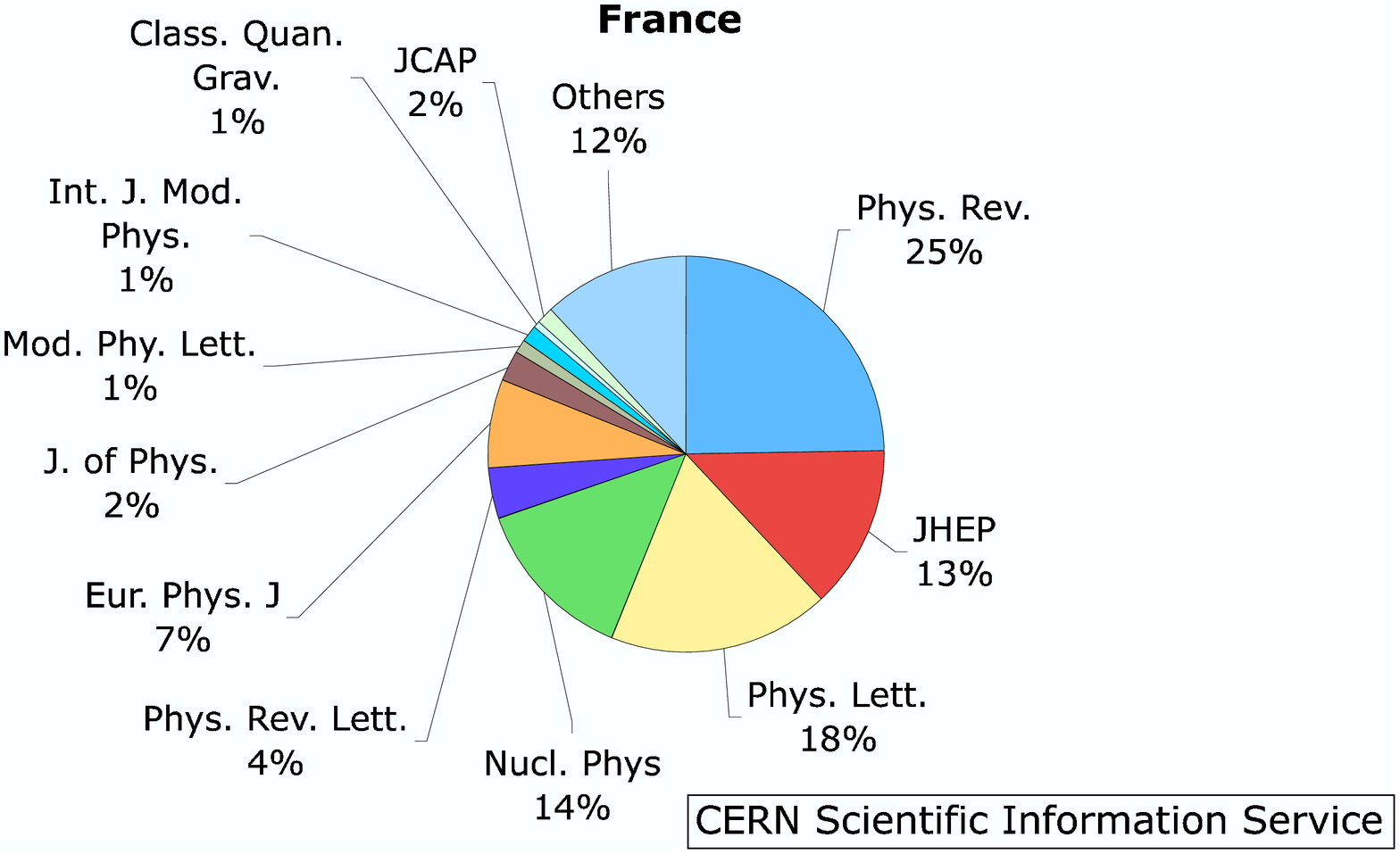}}&
      \mbox{\includegraphics[width=0.5\textwidth]{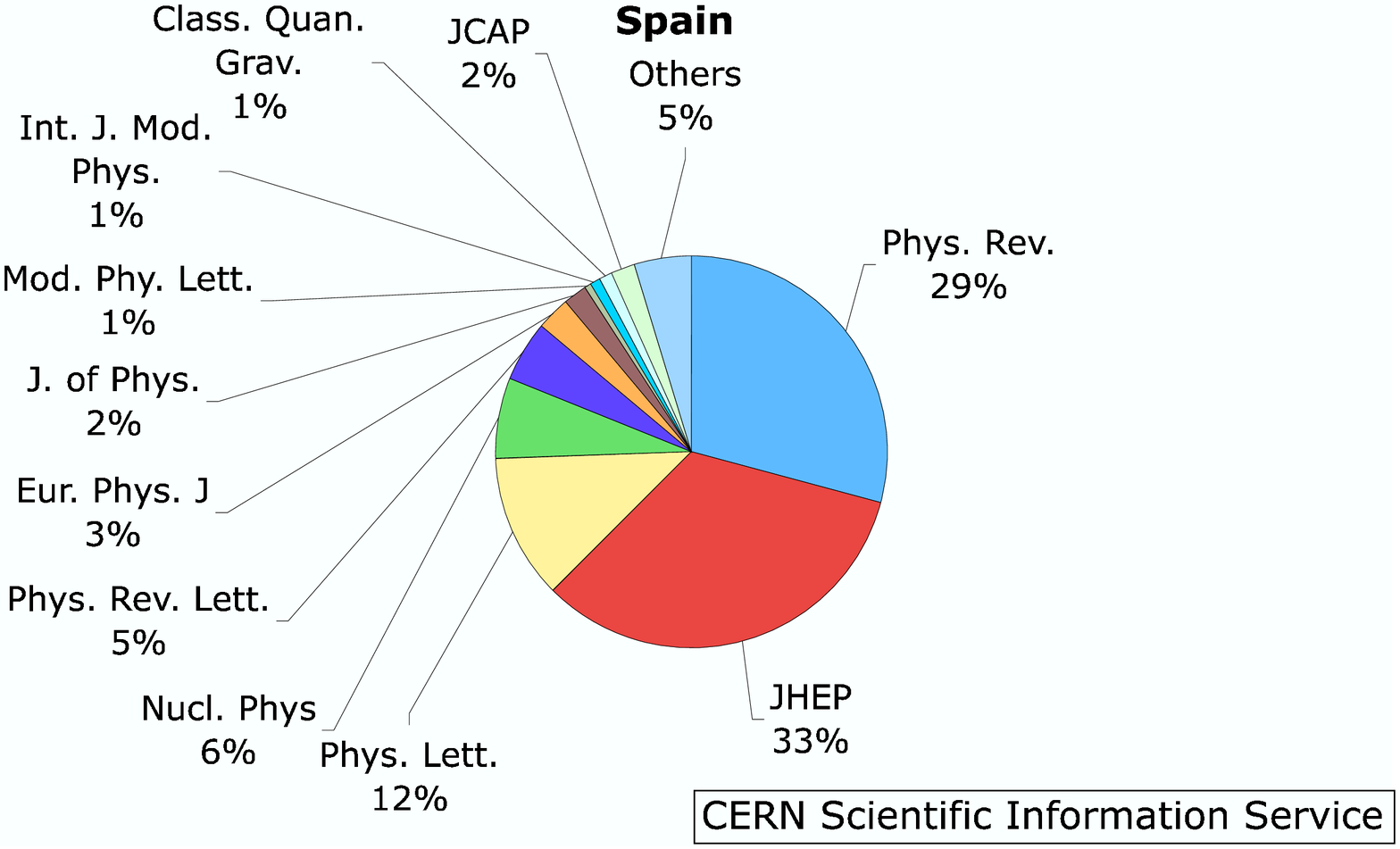}}\\
      \end{tabular}
     \caption{Distribution of HEP articles in different journals
      for several European countries.}
  \end{center}
  \label{fig:9}
\end{figure}\clearpage

\begin{figure}[p]
  \begin{center}
    \begin{tabular}{cc}
      \mbox{\includegraphics[width=0.5\textwidth]{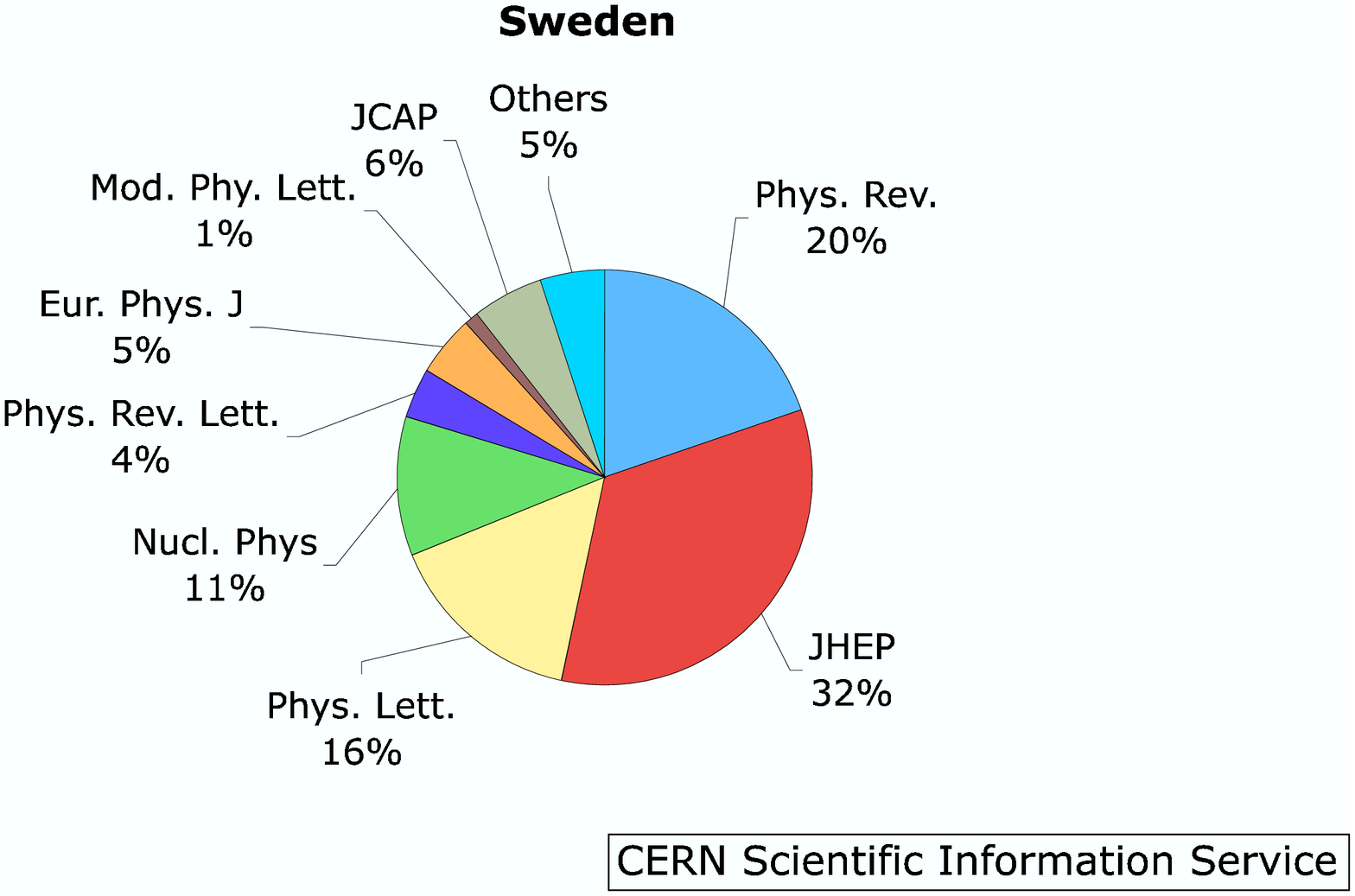}}&
      \mbox{\includegraphics[width=0.5\textwidth]{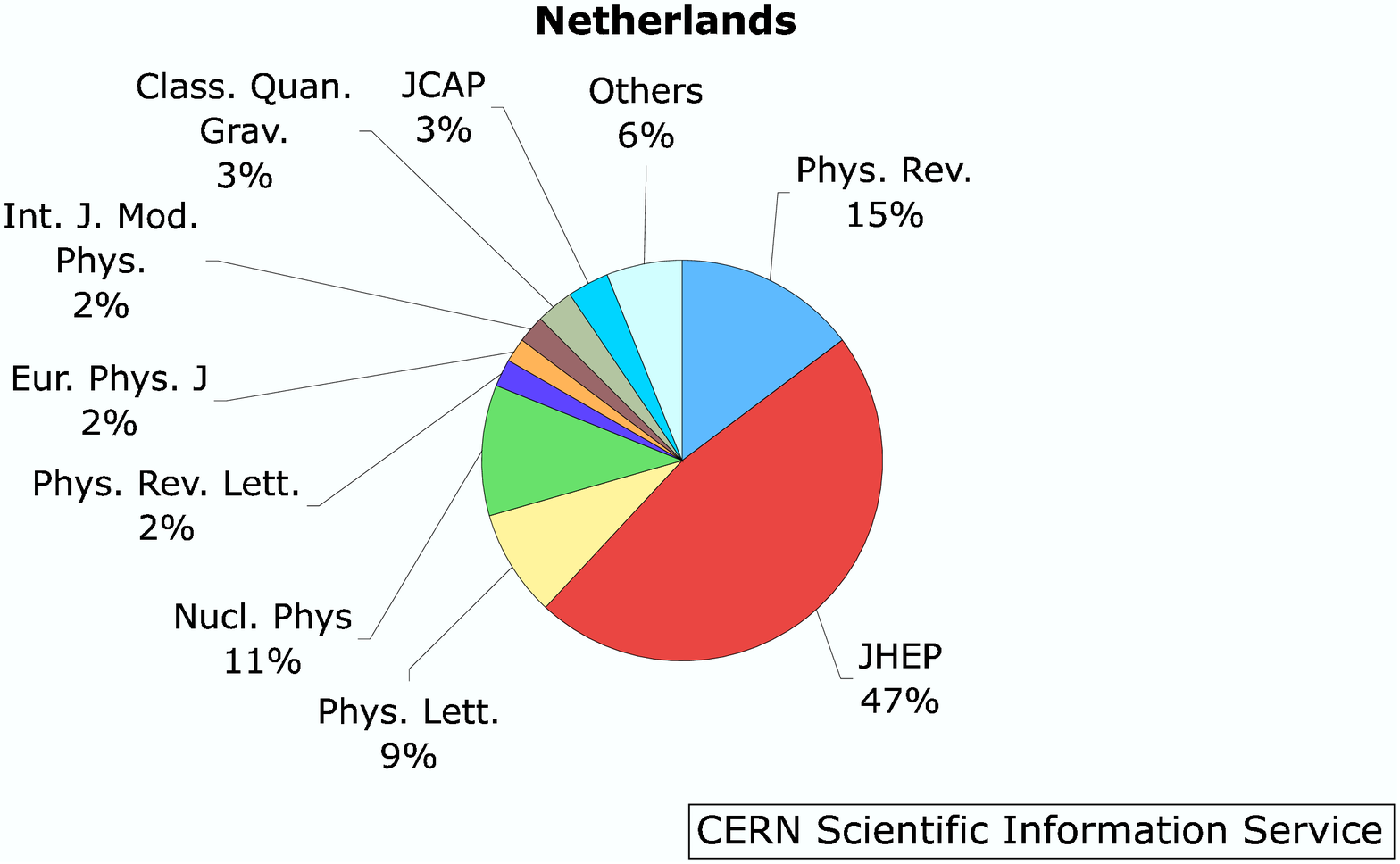}}\\
      \mbox{\includegraphics[width=0.5\textwidth]{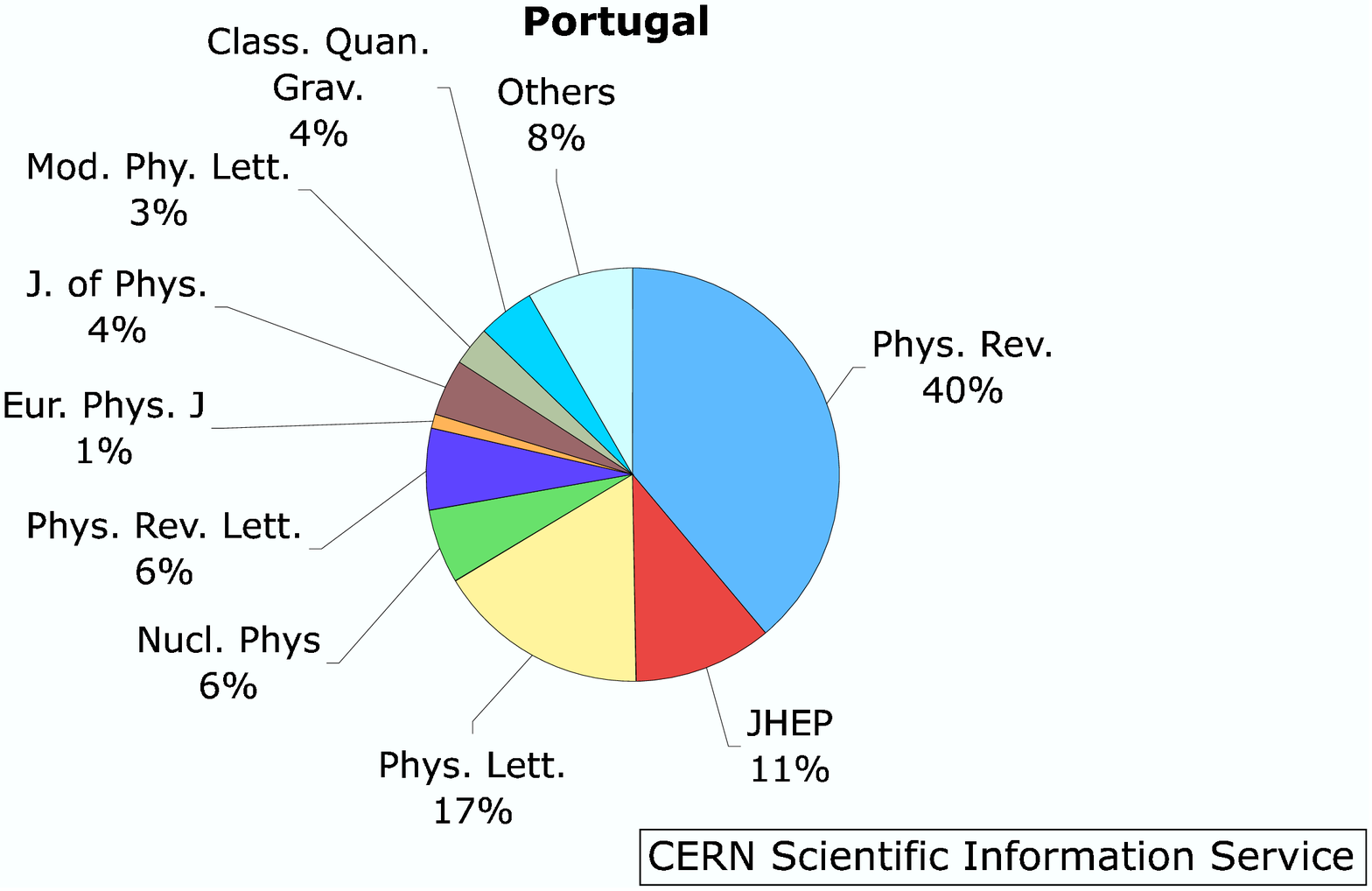}}&
      \mbox{\includegraphics[width=0.5\textwidth]{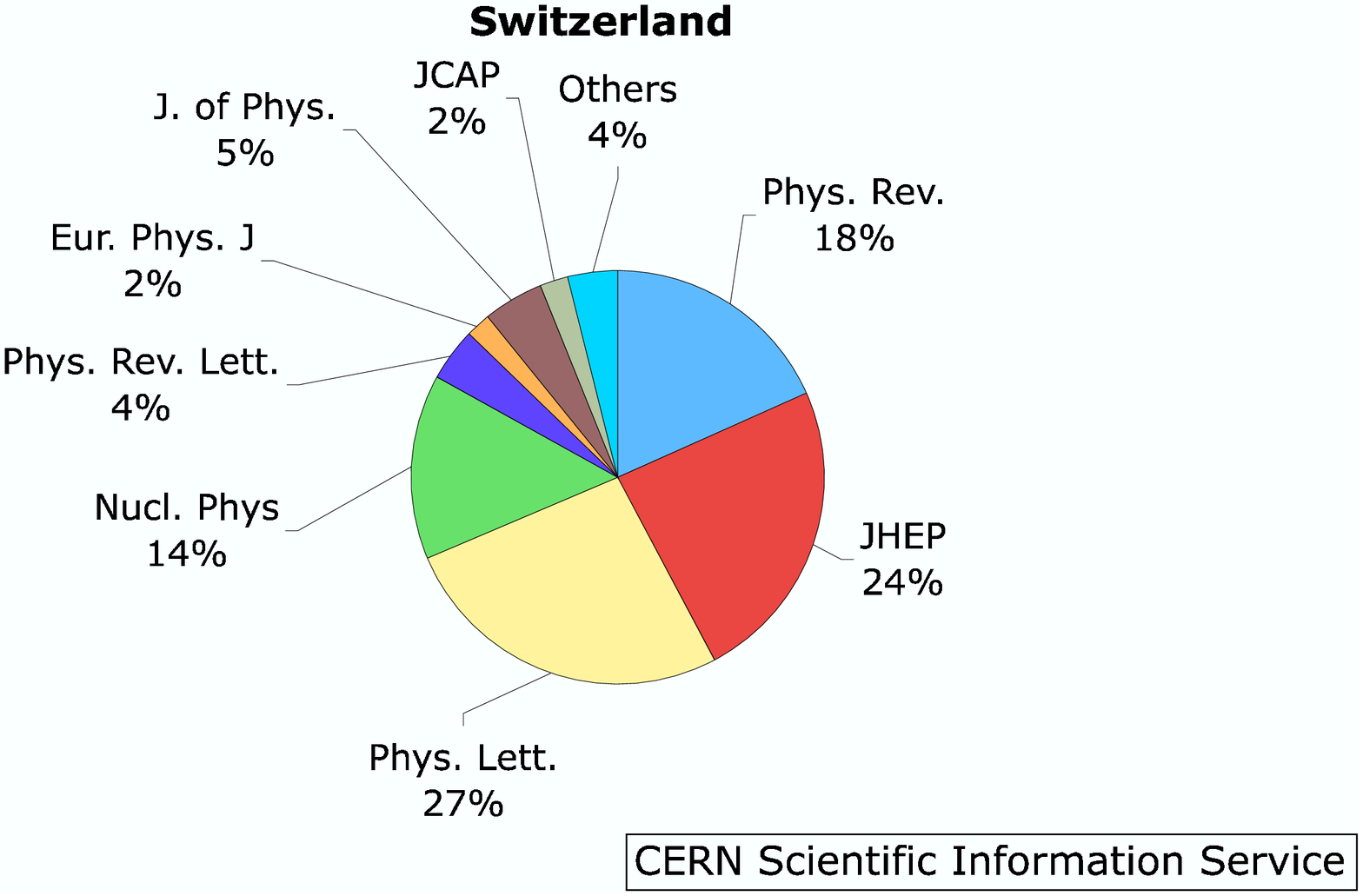}}\\
      \end{tabular}
     \caption{Distribution of HEP articles in different journals
      for several European countries.}
  \end{center}
  \label{fig:10}
\end{figure}\clearpage

\begin{figure}[p]
  \begin{center}
    \begin{tabular}{cc}
      \mbox{\includegraphics[width=0.5\textwidth]{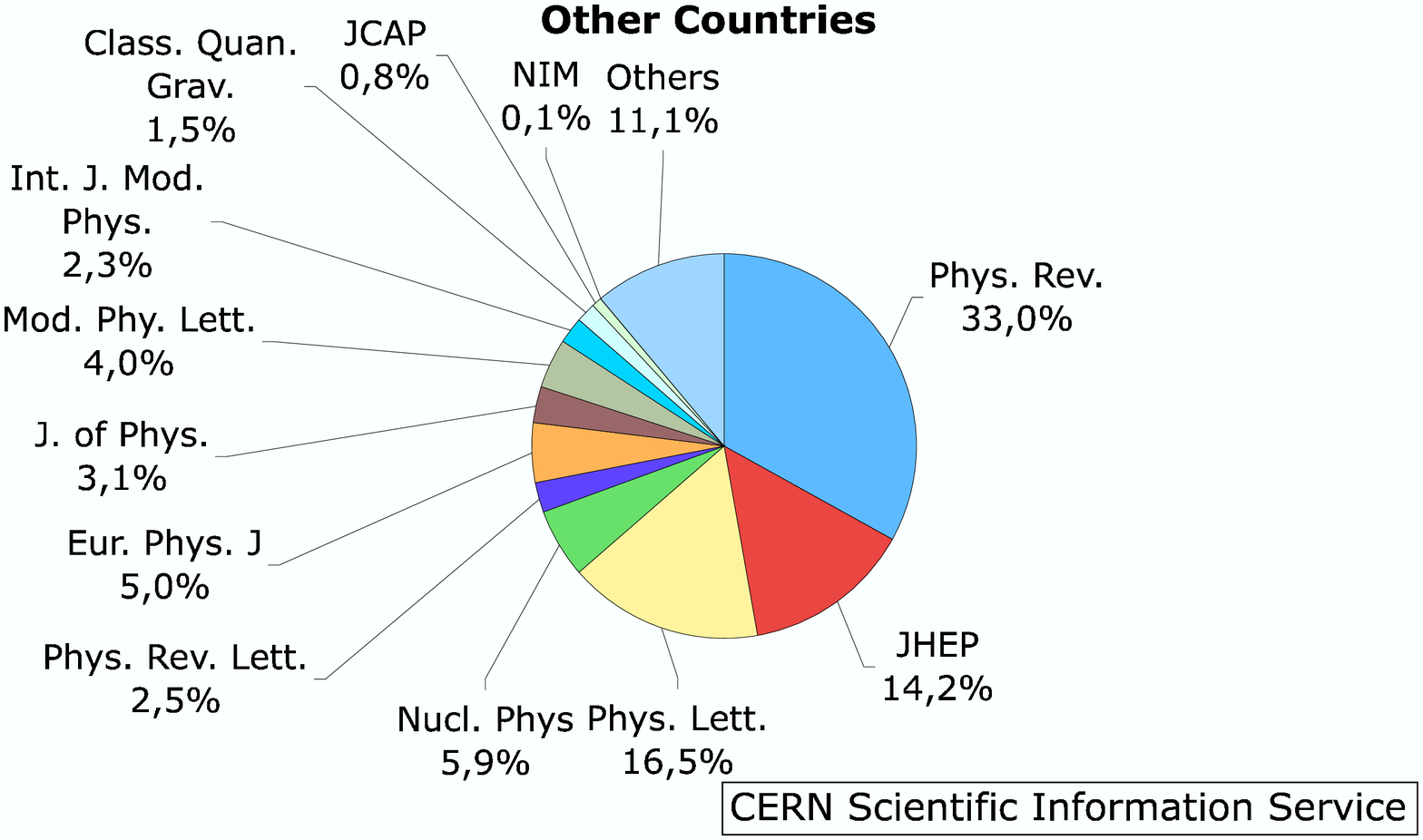}}&
      \mbox{\includegraphics[width=0.5\textwidth]{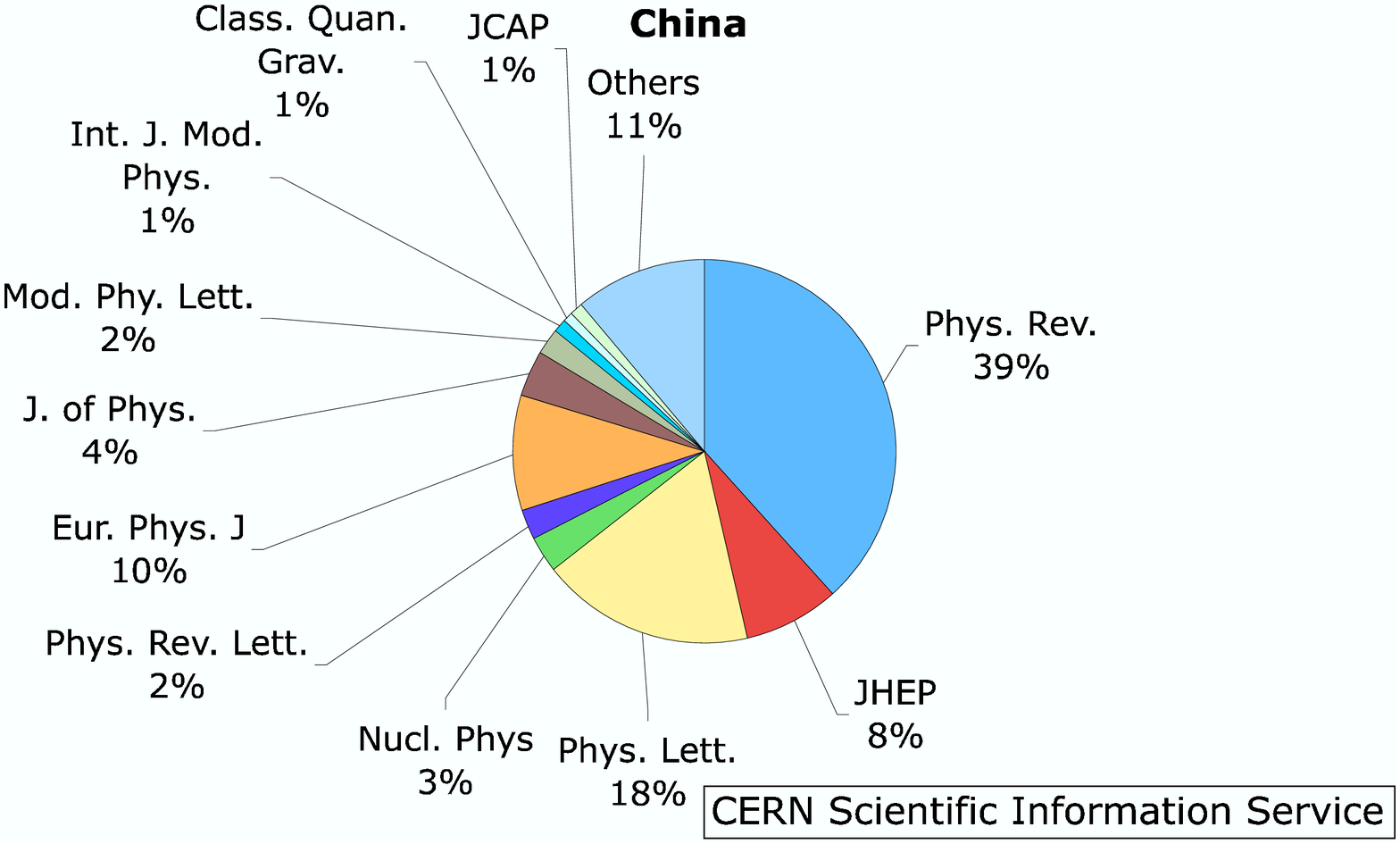}}\\
      \mbox{\includegraphics[width=0.5\textwidth]{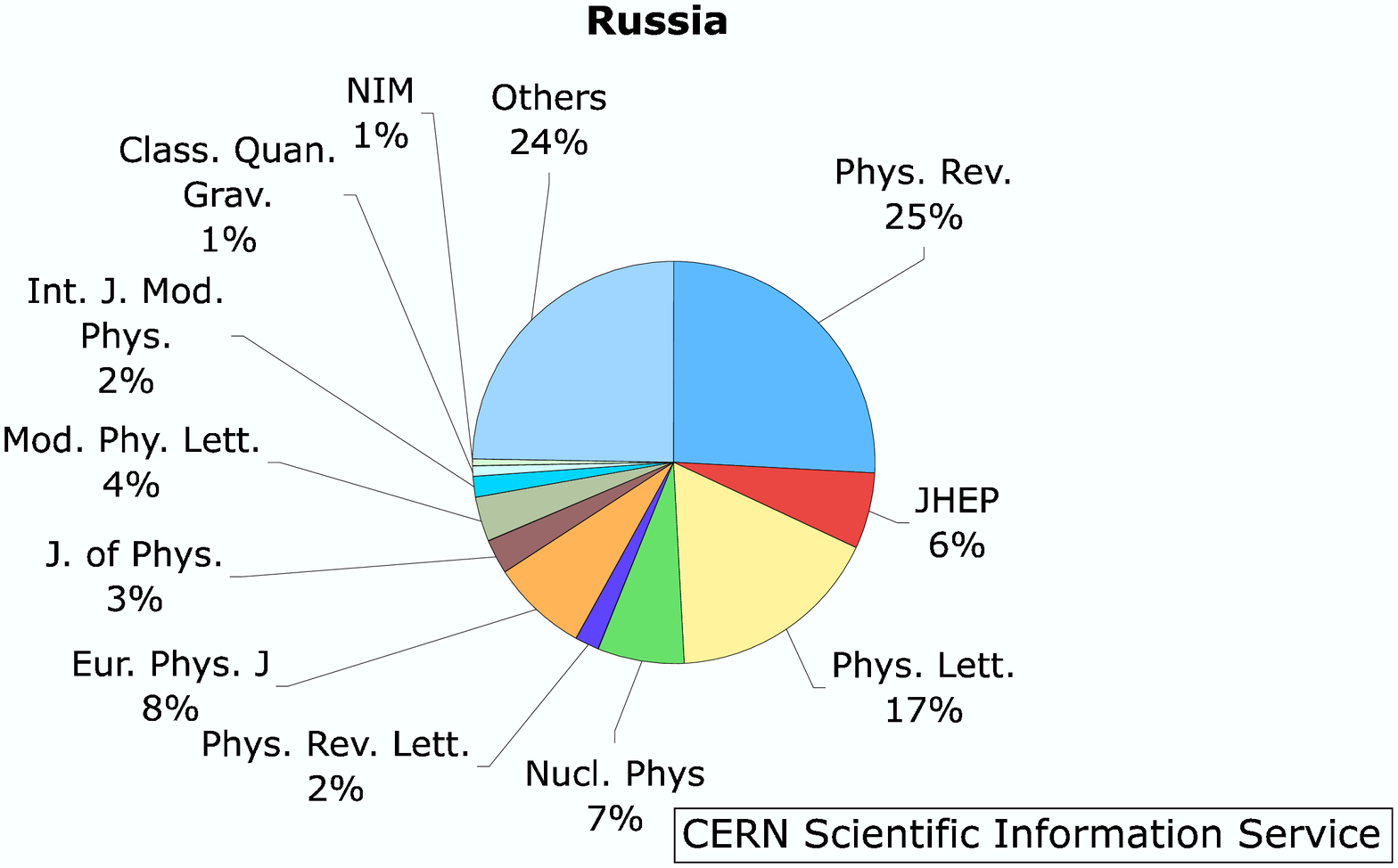}}&
      \mbox{\includegraphics[width=0.5\textwidth]{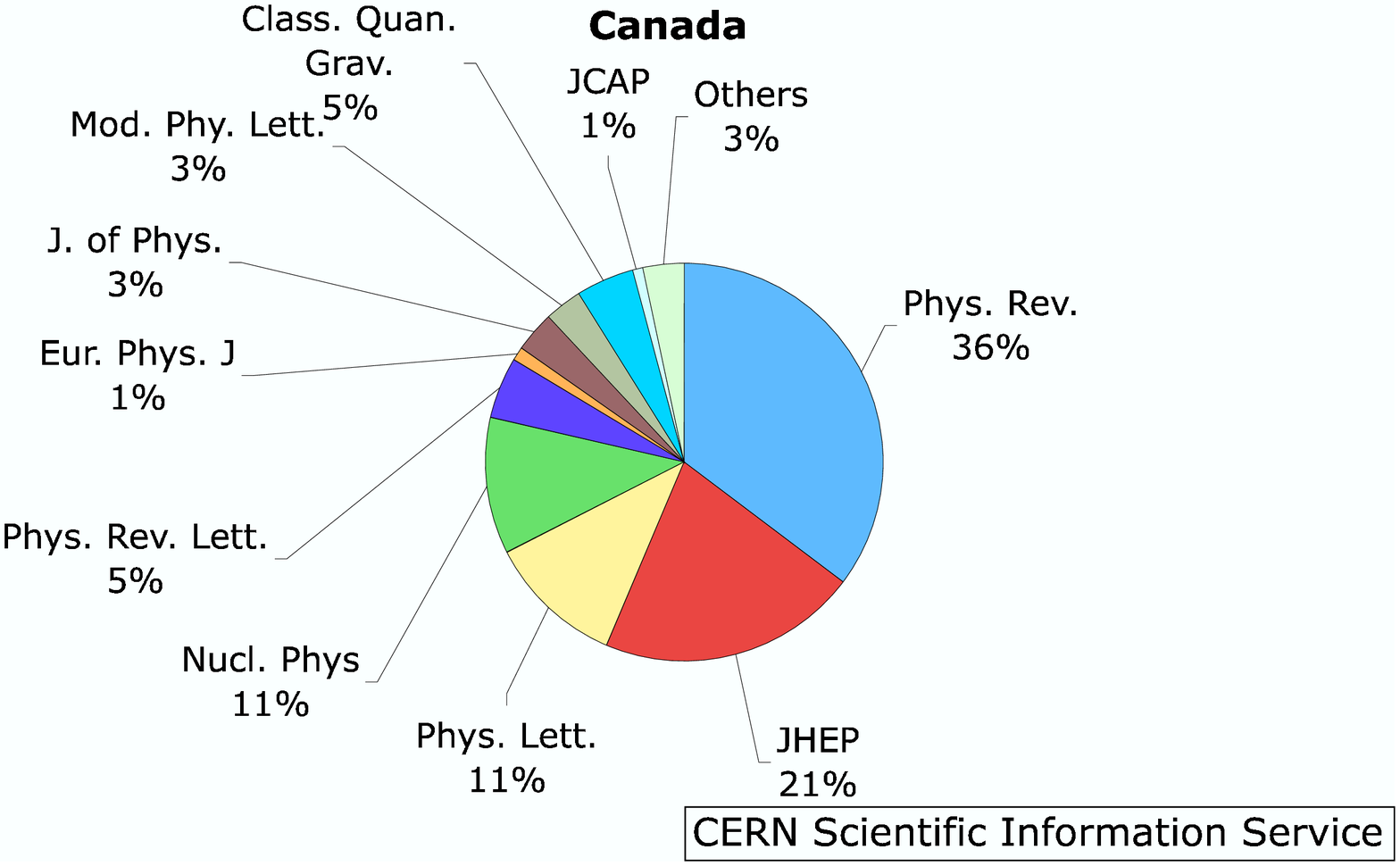}}\\
      \mbox{\includegraphics[width=0.5\textwidth]{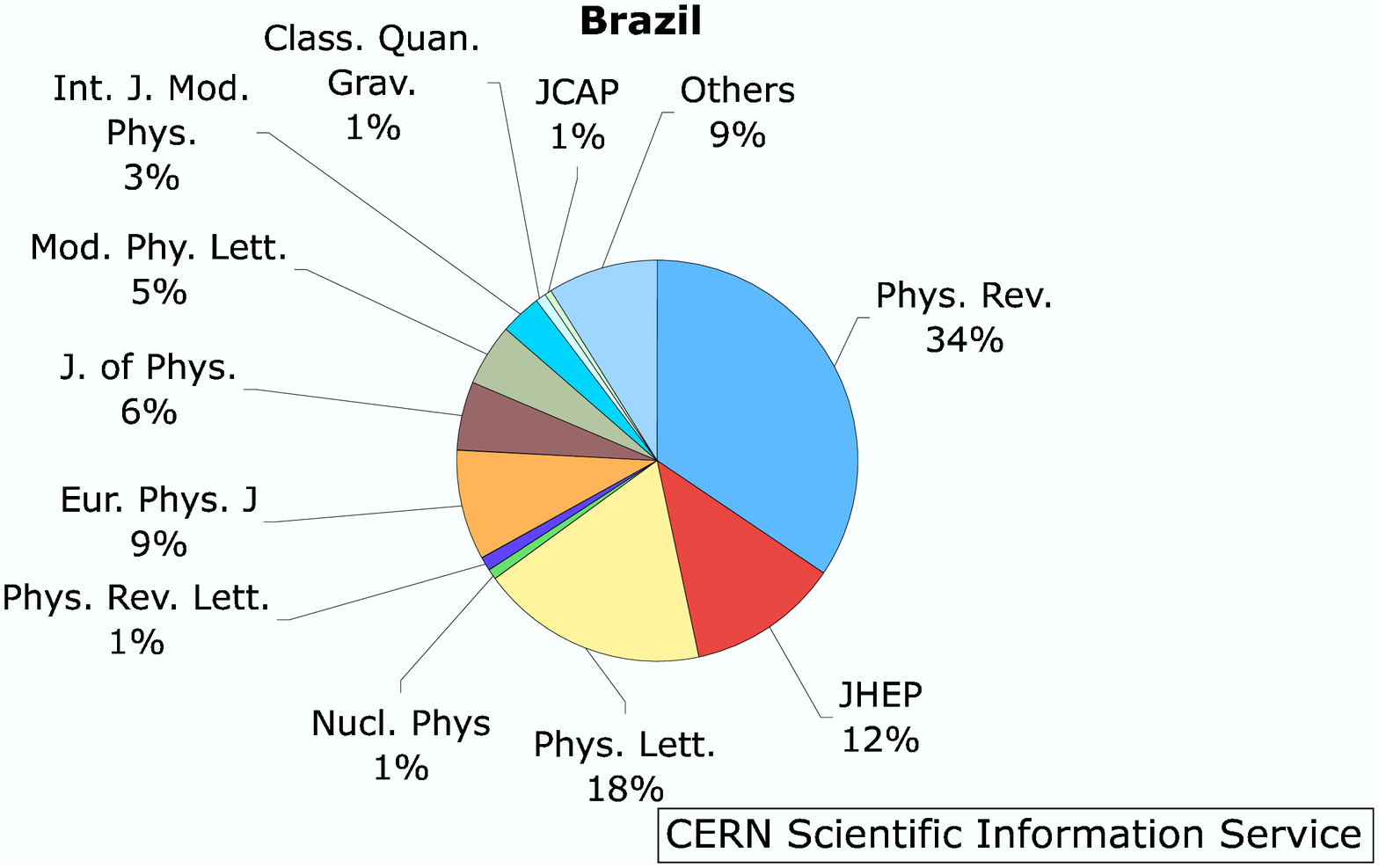}}&
      \mbox{\includegraphics[width=0.5\textwidth]{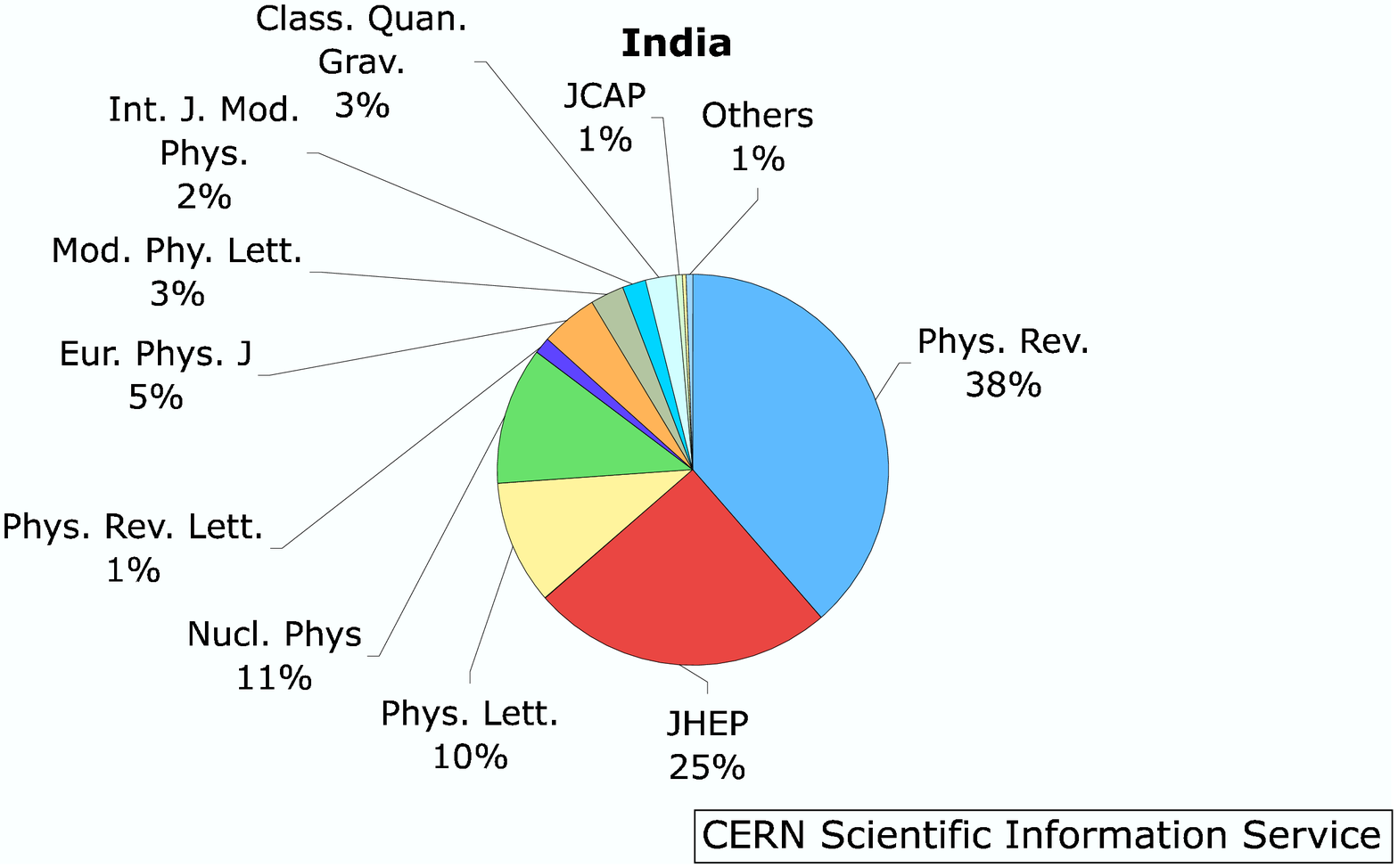}}\\
      \end{tabular}
     \caption{Distribution of HEP articles in different journals
      for several countries.}
  \end{center}
  \label{fig:11}
\end{figure}\clearpage

\end{document}